\documentclass[journal,draftcls,onecolumn,12pt,twoside]{IEEEtranTCOM}

\usepackage{amsmath,amssymb,bm}
\usepackage[dvips]{graphicx}
\usepackage{array}
\usepackage{cite}
\usepackage{enumerate}
\usepackage{color}
\usepackage{float}
\usepackage[capitalise]{cleveref}
\usepackage[mathcal]{euscript}
\usepackage{multirow,multicol,tabu}
\usepackage{subcaption}
\usepackage{soul}

\newcolumntype{?}{!{\vrule width 1pt}}

\usepackage{mathtools}	

\usepackage{algorithm}
\usepackage{algpseudocode}

\usepackage{amsthm}

\newtheorem{remark}{Remark}

\newtheorem{theorem}{Theorem}

\def \Fig {Fig.}

\newcolumntype{M}[1]{>{\centering\arraybackslash}m{#1}}
\newcolumntype{N}{@{}m{0pt}@{}}

\usepackage{etoolbox}

\makeatletter
\newcommand{\changeoperator}[1]{%
  \csletcs{#1@saved}{#1@}%
  \csdef{#1@}{\changed@operator{#1}}%
}
\newcommand{\changed@operator}[1]{%
  \mathop{%
    \mathchoice{\textstyle\csuse{#1@saved}}
               {\csuse{#1@saved}}
               {\csuse{#1@saved}}
               {\csuse{#1@saved}}%
  }%
}
\makeatother

\changeoperator{sum}
\changeoperator{prod}

\usepackage{tikz,tabularx}
\usetikzlibrary{decorations.pathreplacing}
\tikzstyle{every picture}+=[remember picture]

\usetikzlibrary{shapes,arrows,shadows}	
\usepgflibrary{patterns}
\usetikzlibrary{patterns}
\usetikzlibrary{shapes.geometric}
\usetikzlibrary{arrows}

\usetikzlibrary{decorations.markings}

\tikzset{myptr1/.style={decoration={markings,mark=at position 1 with %
    {\arrow[scale=2.5]{>}}},postaction={decorate}}}
    
\tikzset{myptr2/.style={decoration={markings,mark=at position 1 with %
    {\arrow[scale=1.8]{>}}},postaction={decorate}}}
    
\usetikzlibrary{decorations.markings}

\usepackage{multicol}
\usepackage{pgfplots}
\pgfplotsset{major grid style={densely dotted,black!70}}
\pgfplotsset{compat=1.9}

\usepgfplotslibrary{fillbetween}

\pgfplotsset{axis_x_1_y_1_small/.style={
	      width=4.67cm,
	      legend style={at={(0.65,0.5)},anchor=west},
	      xmin=0,xmax=1,
	      ymin=0,ymax=1,
	      grid=both,
	      xtick={0,0.2,0.4,0.6,0.8,1},
 	      ytick={0,0.2,0.4,0.6,0.8,1},	      
	      xticklabels={\scriptsize{0},~,~,~,~,\scriptsize{1}}},
	      yticklabels={~,~,~,~,~,\scriptsize{1}}}	  	       	

\pgfplotsset{axis_x_4_y_1_small/.style={
	      width=4.67cm,
	      legend style={at={(0.65,0.5)},anchor=west},
	      xmin=0,xmax=4,
	      ymin=0,ymax=1,
	      grid=both,
	      xtick={0,0.8,1.6,2.4,3.2,4},
 	      ytick={0,0.2,0.4,0.6,0.8,1},	      
	      xticklabels={\scriptsize{0},~,~,~,~,\scriptsize{4}}},
	      yticklabels={~,~,~,~,~,\scriptsize{1}}}

\pgfplotsset{axis_x_6_y_1_small/.style={
	      width=4.67cm,
	      legend style={at={(0.65,0.5)},anchor=west},
	      xmin=0,xmax=6,
	      ymin=0,ymax=1,
	      grid=both,
	      xtick={0,1.2,2.4,3.6,4.8,6},
 	      ytick={0,0.2,0.4,0.6,0.8,1},	      
	      xticklabels={\scriptsize{0},~,~,~,~,\scriptsize{6}}},
	      yticklabels={~,~,~,~,~,\scriptsize{1}}}	  	      	            
	      
\tikzset{mynode/.style={inner sep=2pt,fill,outer sep=0,circle}}	      
	      
\title{EBP-GEXIT Charts for M-ary AWGN Channel for Generalized LDPC and Turbo Codes}

 \author{
  	\IEEEauthorblockN{Arti Yardi,}
 	\IEEEauthorblockA{IIIT-Bangalore,}
 	\IEEEauthorblockA{arti.yardi@iiitb.ac.in}		
 	\and\\
 	\IEEEauthorblockN{Tarik Benaddi,}
 	\IEEEauthorblockA{IMT Atlantique,}
 	\IEEEauthorblockA{tarik.benaddi@imt-atlantique.fr}
 	\and\\
 	\IEEEauthorblockN{Charly Poulliat,}
 	\IEEEauthorblockA{IRIT/INPT-ENSEEIHT,}
 	\IEEEauthorblockA{charly.poulliat@enseeiht.fr}
 	 \and\\
 	\IEEEauthorblockN{Iryna Andriyanova,}
	\IEEEauthorblockA{ETIS/ENSEA,UCP, CNRS,}
	\IEEEauthorblockA{iryna.andriyanova@ensea.fr}
 }

\begin{document}

\maketitle

%

%
\begin{abstract}
This work proposes a tractable estimation of the maximum a posteriori (MAP) threshold of various families of sparse-graph code ensembles, by using an approximation for the extended belief propagation generalized extrinsic information transfer (EBP-GEXIT) function, first proposed by M{\'e}asson et al. 
We consider the transmission over non-binary complex-input additive white Gaussian noise channel and extend the existing results to obtain an expression for the GEXIT function.
We estimate the MAP threshold by applying the \emph{Maxwell construction} to the obtained approximate EBP-GEXIT charts for various families of low-density parity-check (LDPC), generalized LDPC, doubly generalized LDPC, and serially concatenated turbo codes (SC-TC).
When codewords of SC-TC are modulated using Gray mapping, we also explore where the spatially-coupled belief propagation (BP) threshold is located with respect to the previously computed MAP threshold.
%
Numerical results indicate that the BP threshold of the spatially-coupled SC-TC does saturate to the MAP threshold obtained via EBP-GEXIT chart.
\end{abstract}


%
\section{Introduction}
\label{Section_introduction}
For an arbitrary channel code ensemble, there are three fundamental limits on the channel parameter above which reliable communication is not possible; capacity threshold, MAP threshold, and BP threshold. Capacity threshold corresponds to the limit imposed by the channel, MAP threshold provides the limit for the given specific channel code ensemble\footnote{For example LDPC code ensemble defined by a specific degree distribution for variable and check nodes.}, and BP threshold corresponds to the limit that can be achieved in practice via BP decoding. While finding the capacity and BP thresholds are possible for a various channels and code ensembles, finding the MAP threshold, in general, is known to be difficult~\cite{Urbanke_Richardson_MCT_Book, GAT_Urbanke_2009, Maxell_Urbanke_2008}. 
%

It is known that, multiple copies of a given code ensemble can be appropriately spatially coupled such that the BP threshold of the spatially-coupled code approaches the MAP threshold of the uncoupled ensemble and this phenomenon is termed as \emph{threshold saturation}~\cite{Kudekar_ThresholdSaturationIT}.
Threshold saturation of various families of channel codes has been studied in the literature~\cite{Kudekar_ThresholdSaturationIT, Yedla_potential_func_2014, kumar2014threshold, moloudi2016spatially, SCTC_AWGN_TCOM_2019, Tarik_Globecomm_2017, Yedla_LDPC_BICM_arxiv}.
While threshold saturation phenomenon has been proved analytically for the binary erasure channel (BEC), it is conjectured for numerically involved codes such as generalized LDPC (GLDPC) and serially concatenated turbo codes (SC-TC) over the additive white Gaussian noise (AWGN) channel.
%
%
Since the MAP threshold is the fundamental limit that one can hope to approach via spatial coupling, it is desirable to find the MAP threshold of the uncoupled code ensemble. 
M{\'e}asson et al.~have proposed an analytical method to estimate the MAP threshold of LDPC and parallel turbo codes over any binary memoryless symmetric (BMS) channel.
In this method, the MAP threshold is obtained by applying the \textit{Maxwell construction} to the EBP-GEXIT chart of the given code ensemble~(details can be found in \cite[Sec.~3.20]{Urbanke_Richardson_MCT_Book}, \cite{Maxell_Urbanke_2008}).
For the BEC, when the closed form expression for the density evolution (DE) equation is known, the EBP-GEXIT chart can be obtained analytically~\cite{Maxell_Urbanke_2008}. 
While finding this EBP-GEXIT chart is possible for the BEC, obtaining it for general BMS may become computationally prohibitive for numerically involved codes such as GLDPC and doubly generalized LDPC (DGLDPC) codes. 
In \cite{Our_ISIT_2018} and \cite{Our_ISIT_2019}, authors have proposed a method to find an approximate EBP-GEXIT chart for LDPC/GLDPC/DGLDPC codes and SC-TC respectively.
The existing works, either for obtaining the exact or approximate EBP-GEXIT charts, consider the situation when the underlying channel is BMS. 
Obtaining the EBP-GEXIT charts for a complex-input non-binary channel is also desirable.

%
In this paper, we study the problem of finding an approximate EBP-GEXIT chart for ensembles of LDPC/GLDPC/DGLDPC codes and SC-TC, when the transmission is over non-binary complex-input AWGN channel
and the codewords are modulated by an arbitrary modulation scheme.
We first consider the situation when the constellation of modulated symbols is mapped according the Gray mapping.
In this case, we consider the equivalent bit-interleaved coded-modulation (BICM) channel~\cite{BICM_Fabregas_Book}, which is a
BMS channel and provide a method to find an approximate EBP-GEXIT chart.
For the situation of an arbitrary mapping where the channel inputs could be complex alphabets, we extend the existing results to obtain an expression for the GEXIT function for complex AWGN channel and then provide a numerical approximation method to find the EBP-GEXIT chart.
For all the cases, we estimate the MAP thresholds for various codes by applying the Maxwell construction to these approximate EBP-GEXIT charts.
When codewords of SC-TC are mapped according to Gray mapping, 
we also explore where the spatially-coupled BP threshold is located with respect to the previously estimated MAP threshold by using the formalism provided in \cite{Tarik_Globecomm_2017}.
We observe that the BP threshold of the spatially-coupled SC-TC does saturate to the MAP threshold obtained via our approximate EBP-GEXIT chart.

\textit{Organization:} In \cref{Section_System_Model}, we discuss the system model and summarize some preliminaries about the EBP-GEXIT charts.
The proposed numerical method for tractable computation of an approximate EBP-GEXIT chart for Gray and non-Gray mapping is discussed in \cref{Section_EBP_Gray} and \cref{Section_EBP_non_Gray} respectively.
Spatial coupling of SC-TC is discussed in \cref{Section_Spatial_Coupling}.
In \cref{Section_Simulations}, we provide numerical results for several LDPC/GLDPC/DGLDPC and serially concatenated turbo codes
and finally conclude in \cref{Section_Conclusion}. 

%
%
\section{System model and preliminaries}
\label{Section_System_Model}

\subsection{System model}

LDPC$(\lambda, \rho)$ denotes LDPC code ensemble, where 
$\lambda(x)=\sum_i \lambda_i x^{i-1}$ and $\rho(x)=\sum_j \rho_j x^{j-1}$ denote the edge perspective degree distributions for variable nodes (VNs) and check nodes (CNs) respectively~\cite{Urbanke_Richardson_MCT_Book}.
Let $\Lambda(x)$ and $P(x)$ be the corresponding node perspective degree distribution pairs for VNs and CNs respectively. 
While for LDPC codes, every CN correspond to the single parity check code and every VN correspond to the repetition code, 
for GLDPC codes some of the CNs correspond to an arbitrary linear code and 
for DGLDPC codes both VNs and CNs correspond to a general linear code~\cite{Ryan_Costello_Book}. We assume that all VNs are unpunctured and have degrees strictly greater than one.
%
%
For SC-TC, we denote the outer and inner convolutional codes by $\mathcal{O}$ and $\mathcal{I}$ respectively and the corresponding SC-TC code ensemble is denoted by $\mathcal{S}(\mathcal{O},\mathcal{I})$. 
%

%
%
\begin{figure*}[t]

\begin{center}

\begin {tikzpicture}

	  \draw [->](-0.45,0) -- (0,0);     
	  \node [above] at (-0.25,0) {$\mathbf{b}$};	  	  
    
	  \draw [rounded corners] (0,-0.5) rectangle (1.6,0.5);      
 	  \node [right] at (-0.03,0.2) {Channel};	  	  
 	  \node [right] at (0.15,-0.2) {code};	  	   	  
	  
	  \draw [->](1.6,0) -- (1.6+0.5,0);     
	  \node [above] at (1.6+0.2,0) {$\mathbf{c}$};	  	  
  
	  \draw [rounded corners] (1.6+0.5,-0.5) rectangle (1.6+0.5+1.9,0.5);      
 	  \node [right] at (1.6+0.4,0) {Interleaver};	  	  

	  \draw [->](1.5+0.5+1.7+0.3,0) -- (1.5+0.5+1.7+0.5+0.3,0);     
	  \node [above] at (1.5+0.3+1.7+0.5+0.3,0) {$\mathbf{c}^{\prime}$};	  	  
	  
	  \draw [rounded corners] (1.5+0.5+1.7+0.5+0.3,-0.5) rectangle (1.5+0.5+1.7+0.5+1.7+0.65,0.5);      
 	  \node [right] at (1.5+0.5+1.7+0.5+0.3,0.22) {Modulator};	  	  	  
 	  \node [right] at (1.5+0.5+1.7+0.5+0.3,-0.22) {+ Mapper};	  	  	   	  
 	  
 	  \draw [->](1.5+0.5+1.7+0.5+1.7+0.65,0) -- (1.5+0.5+1.7+0.5+1.7+0.5+0.65,0);     
	  \node [above] at (1.5+0.5+1.7+0.5+1.7+0.25+0.65,0) {$\mathbf{x}$};	  	   	  

	  \draw [rounded corners] (1.5+0.5+1.7+0.5+1.7+0.5+0.65,-0.5) rectangle (1.5+0.5+1.7+0.5+1.7+0.5+1+0.5+0.7,0.5);      
	  \node [right] at (1.5+0.5+1.7+0.5+1.7+0.53+0.55,0) {Channel};	  
 	  
 	  \draw [->](7.4+0.5+0.7,0) -- (7.4+0.5+0.5+0.7,0);     	   	  
 	  \node [above] at (7.4+0.25+0.5+0.7,0) {$\mathbf{y}$};	  	   	  
 	  \draw [rounded corners] (7.4+0.5+0.5+0.7,-0.5) rectangle (7.4+0.5+2.1+0.7,0.5);       	  
  	  \node [right] at (7.4+0.5+0.55+0.6,0) {Detector};	  	  	  
 	  
 	  \draw [->](7.4+0.5+2.1+0.7,0) -- (7.4+0.5+2.1+0.5+0.7,0);   	   	  
 	  \node [above] at (7.4+0.5+2.1+0.18+0.75,0) {$\mathbf{L}^{\prime}$};	  	   	  
 	  
	  \draw [rounded corners] (7.4+0.5+2.1+0.5+0.7,-0.5) rectangle (7.4+0.5+2.1+0.5+2.1+1,0.5);       	  
	  
 	  \node [right] at (7.4+0.5+2.1+0.5+0.6,0) {Deinterleaver};	  	  	  
 	  
 	  \draw [->](7.4+0.5+2.1+0.5+2.1+1,0) -- (7.4+0.5+2.1+0.5+2.1+0.5+1,0);     	   	   	   	  
 	  \node [above] at (7.4+0.5+2.1+0.5+2.1+0.3+1,0) {$\mathbf{L}$};	  	   	  
	  
	  \draw [rounded corners] (7.4+0.5+2.1+0.5+2.1+0.5+1,-0.5) rectangle (7.4+0.5+2.1+0.5+2.1+0.5+1.5+1.1,0.5);      
 	  \node [right] at (7.4+0.5+2.1+0.5+2.1+0.5+1,0.2) {Channel};	  	  
 	  \node [right] at (7.4+0.5+2.1+0.5+2.1+0.5+1,-0.2) {decoder};	  	   	   	  
 	  
 	  \draw [->](7.4+0.5+2.1+0.5+2.1+0.5+1.5+1.1,0) -- (7.4+0.5+2.1+0.5+2.1+0.5+1.5+0.5+1,0);     	   	   	   	   	  
 	  \node [above] at (7.4+0.5+2.1+0.5+2.1+0.5+1.5+0.25+1.05,0) {$\hat{\mathbf{b}}$};	  	   	  
 	  
  	  \draw [rounded corners, dashed, thick] (3.82-1.85,-1) rectangle (10.8-0.5+2.4+1,1);      
   	  \draw [->](10.8-0.5+2.4+1,-1) -- (11.2-0.5+2.35+1,-1.2);    
   	  \node [right] at (11.2-0.5+2.35+1,-1.27+0.15) {BICM};	
   	  \node [right] at (11.2-0.5+2.35+1,-1.27-0.15) {channel};

\end{tikzpicture}

\end{center}
\vspace{-0.5cm}
\caption{Block diagram of a digital communication system considered in this paper.}
\vspace{-0.8cm}
\label{Figure_general_digi_comm_system}  	  
\end{figure*}
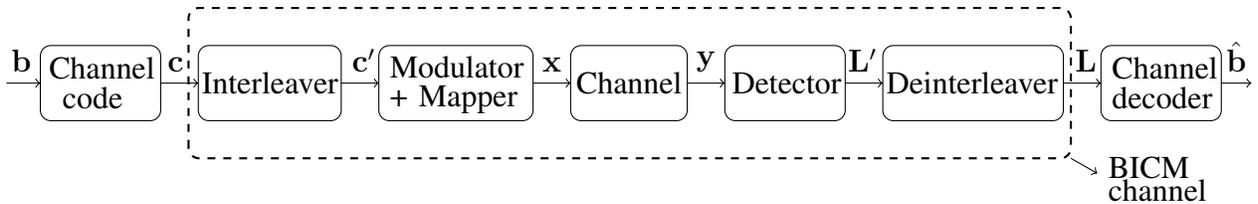
%
We consider the digital communication system illustrated in \cref{Figure_general_digi_comm_system}. 
The binary channel code used at the transmitter is either a LDPC code or SC-TC.
A sequence of message bits $\mathbf{b}$ is encoded to get a codeword sequence $\mathbf{c} = [ c_1 \mbox{~} c_2 \mbox{~} \ldots \mbox{~} c_n]$ of length $n$.
The encoded bits are then interleaved to get an interleaved sequence of codebits $\mathbf{c}^{\prime} = [ c_1^{\prime} \mbox{~} c_2^{\prime} \mbox{~} \ldots \mbox{~} c_n^{\prime}]$. 
For a $2^m$-ary modulation scheme, coded bit sequence $\mathbf{c}^{\prime}$ is first divided into $N = n/m$ vectors each of length $m$ and denoted by 
$\mathbf{c}^{\prime} = [ \mathbf{c}^{\prime}(1) \mbox{~} \mathbf{c}^{\prime}(2) \mbox{~} \ldots \mbox{~} \mathbf{c}^{\prime}(N)]$, where 
$\mathbf{c}^{\prime}(t) = [ c_{(t-1)m+1}^{\prime} 
\mbox{~} c_{(t-1)m+2}^{\prime} \mbox{~} \ldots \mbox{~} c_{tm}^{\prime} ]$ for $t = 1,2, \ldots, N$.
Let us denote the entries in $\mathbf{c}^{\prime}(t)$ as $\mathbf{c}^{\prime}(t) = [ c_{t,1}^{\prime} \mbox{~} c_{t,2}^{\prime} \mbox{~} \ldots \mbox{~} c_{t,m}^{\prime}]$.
Modulated symbols are denoted by $\mathbf{x} = [ x_1 \mbox{~} x_2 \mbox{~} \ldots \mbox{~} x_N ]$, where each $x_t$ is a complex modulated symbol corresponding to $\mathbf{c}^{\prime}(t)$. 
Let $\mathbb{X} = \{ \xi_1, \xi_2, \ldots, \xi_{|\mathbb{X}|} \}$ denotes the set of complex constellation symbols.
The constellation of modulated symbols can be labeled according to any arbitrary mapping rule.
%

The noise is introduced by the AWGN channel with noise variance $\sigma^2$. 
The noise affected version $y_t$ of the $t$-th transmitted symbol $x_t$ is given by $y_t = x_t + n_t$,
where each noise sample $n_t$ is chosen independently and identically distributed (i.i.d.) according to the complex normal distribution $\mathcal{CN}(0,\sigma^2)$, for $t = 1,2,\ldots, N$.
%
%
In this work, it is convenient to parameterize the channel using its entropy, denoted by $h = H(X)$, ($h \in [0,|\mathbb{X}|]$)~\cite{BICM_Fabregas_Book}.
%
%
The family of AWGN channels parameterized by entropy $h$ is denoted by $\{\mbox{AWGN}(h)\}_{h}$ and as in \cite{Urbanke_Richardson_MCT_Book, GAT_Urbanke_2009},
we assume that $\{\mbox{AWGN}(h)\}_{h}$ is ordered by physical degradation and it is smooth with respect to $h$.

At the receiver, the detector computes the sequence $\mathbf{L}^{\prime}$ of log-likelihood ratios (LLR) of the interleaved bit sequence $\mathbf{c}^{\prime}$ and $i$-th entry in $\mathbf{L}^{\prime}$ is given by $L_i^{\prime} \coloneqq \log \frac{\mathbb{P}[c_i^{\prime} = 0|\mathbf{y}]}{\mathbb{P}[c_i^{\prime} = 1|\mathbf{y}]}$, for $i = 1, 2, \ldots, n$. 
The deinterleaved LLR sequence $\mathbf{L}$ is then given to the corresponding channel decoder.
Note that in practice, iterative schemes are considered within the channel decoder (in the case of LDPC or SC-TC) or between the detector and the decoder (in the case of Bit interleaved coded modulation (BICM)). To keep the system model general, we shall discuss the details about the iterative strategies in the corresponding sections.

%

For BMS channels, the input alphabet set is given by $\mathbb{X} = \{+1,-1\}$ and
in this case, the distribution of $\mathbf{L}$ under the condition $X=+1$ is referred to as $L$-density, denoted by $c_{BMS(h)}$~\cite[Sec.~II]{GAT_Urbanke_2009}.
For the binary-input AWGN (BAWGN) channel, the $L$-density is given by $c_{BAWGN(h)} = \mathcal{N}(2/\sigma^2, 4/\sigma^2)$~\cite[Ex.~4.21]{Urbanke_Richardson_MCT_Book} and its entropy $H(c_{BAWGN(h)})$ is given by~\cite[Sec.~II]{GAT_Urbanke_2009},
\begin{equation}
\begin{aligned}
%
H(c_{BAWGN(h)}) = \mathbb{E}_L\big[\log_2(1+e^{-L})\big]
 \coloneqq 1 - J(2/\sigma),
\end{aligned}
\label{Eqn_BAWGN_Entropy}
\end{equation}
where $\mathbb{E}_L[.]$ denotes the expectation with respect to the random variable $L$ and the function $J(.)$ is defined in \cite{Brink_EXIT_2004}. Note that $J(.)$ a function of the standard deviation $2/\sigma$ of the $L$-density $c_{BAWGN(h)}$.
%
%
Channel with non-binary inputs is known to be asymmetric and $L$-density cannot be defined in this case~\cite{Bennatan_Non_binary}.
The entropy of non-binary complex input channel is given by~\cite{Pfister_ISI_2012}
%
\begin{equation}
\begin{aligned}
h = H(X) - \int_{y} \sum_{x \in \mathbb{X}} p(x) p(y|x) \log_2\left\{ \frac{p(y|x)}{p(y)} \right\} \mathrm{d}y.
\end{aligned}
\label{Eqn_NBAWGN_Entropy}
\end{equation}
%

%
%

\subsection{EBP-GEXIT charts of LDPC codes over BAWGN channel \cite{GAT_Urbanke_2009}}
\label{subsection_EBP_GEXIT_preliminaries}
%
%
Let us first recall DE equations for the BP decoding of LDPC codes.
%
%
%
%
The DE equations for LDPC($\lambda, \rho$) code ensemble are given by \cite[Sec.~4.1.4]{Urbanke_Richardson_MCT_Book}
\begin{align}
a^{BP,l} = c_{BAWGN(h)} \star \lambda(\rho(a^{BP,l-1})),
\label{Eqn_DE_general_LDPC}
\end{align}
where $a^{BP,l}$ be the density of a randomly chosen VN to CN message in the $l$-th iteration of BP decoding and $\star$ is the convolution operator.
%
%
For defining the EBP-GEXIT chart, we need to consider a \textit{complete fixed-point (FP) family}~\cite[Sec.~VII-A]{GAT_Urbanke_2009}. The family of densities $\{a_x\}_x$ and $\{c_x\}_x$ parameterized by $x \in [0,1]$ is called a complete FP family if (i) $c_x \in \{ \mbox{BAWGN}(h)\}_h$ for some $h \in [0,1]$, (ii) for any $x \in [0,1]$, $a_x = c_x \star \lambda(\rho(a_x))$ ($a_x$ is a FP density with respect to $c_x$),
(iii) $H(a_x) = x$, and (iv) $\{a_x\}_x$ and $\{c_x\}_x$ are smooth with respect to $x$.
The EBP-GEXIT function, denoted by $g^{EBP}(x)$, for LDPC($\lambda, \rho$) code ensemble is then defined as
\begin{align} 
g^{EBP}(x) \coloneqq \int_{-\infty}^{\infty}  \Lambda(\rho(a_x))(z) l^{c_{x}}(z) \mathrm{d}z,	
\label{Eqn_EBP_GEXIT_func}
\end{align}
where $l^{c_{x}}(z)$ is called the GEXIT kernel~\cite[Ch.4]{Urbanke_Richardson_MCT_Book}. 
%
For the BAWGN channel with $L$-density $c_{BAWGN(h)}=\mathcal{N}(2/\sigma^2,2/\sigma^2)$, an expression for $l^{c_{BAWGN(h)}}(z)$ is given by~\cite[Example~7]{GAT_Urbanke_2009}
\begin{equation}
\begin{aligned}
& l^{c_{BAWGN(h)}}(z) = 
&\left( \int_{-\infty}^{\infty} \frac{e^{- \frac{(w-(2/\sigma^2))^2}{8/ \sigma^2}}}{1+e^{w+z}} \mbox{d}w \right) \bigg/ \left( \int_{-\infty}^{\infty} \frac{e^{- \frac{(w-(2/\sigma^2))^2}{8/ \sigma^2}}}{1+e^{w}} \mbox{d}w \right). 
\end{aligned}
\label{Eqn_BAWGN_kernel}
\end{equation}
The EBP-GEXIT chart is the curve obtained by plotting $g^{EBP}(x)$ versus $c_x$ $\forall x \in [0,1]$.
%

%
\section{EBP-GEXIT chart over the AWGN channel with Gray mapping}
\label{Section_EBP_Gray}

In this section, we consider the situation when modulated symbols are mapped according to the Gray mapping and study the situation of any arbitrary mapping in the next section. 

%
\subsection{Equivalent bit-channels for Gray mapping}
\label{subSection_BICM channel}

In the presence of an interleaver between the channel coded bits and the modulator, in \Fig~\ref{Figure_general_digi_comm_system} one can consider an equivalent channel, termed as the \textit{BICM channel}, formed by the interleaver, modulator, AWGN channel, detector, and deinterleaver~\cite{BICM_Fabregas_Book, BICM_Szczecinski_book}.
%
%
For obtaining the EBP-GEXIT chart, we parameterize this BICM channel by its entropy $h$, denoted as BICM$(h)$.
An ideal interleaver implies that the set of random variables corresponding to $\{ L_1, L_2, \ldots, L_n \}$ are independent and hence this BICM channel can be equivalently seen as a set of $n$ parallel independent BMS~\cite{i2008bit}.
%
%
For some modulation schemes, the exact distribution of each $L_i$ is known. For example, for BPSK modulation with Gray mapping, each $L_i$ is i.i.d.~according to $\mathcal{N}(2/\sigma^2, 4/\sigma^2)$. 
However for an arbitrary modulation scheme, the distribution of LLRs need not be Gaussian or identical.
Further, finding this distribution may become difficult
and hence in the literature good approximations for its distribution are suggested\footnote{For example, F\`{a}bregas et al.~have suggested a Gaussian approximation for the Gray labeling~\cite{Caire_GA}. For PSK and rectangular QAM, Szczeci{\'n}ski et al.~have suggested approximation as a mixture of Gaussian functions in \cite{Szczecinski_PSK} and \cite{Szczecinski_QAM} respectively.}.

Typically for the Gray mapping, the density of a randomly chosen $L_i$ can be well approximated by a mixture of consistent Gaussian densities.
%
%
For an integer $M$ suppose $c_{BICM(h)}$ is given by
\begin{align}
c_{BICM(h)} = \sum_{j = 1}^M d_j \mathcal{N} (l_j, 2l_j),
\label{Eqn_GA_mixture}
\end{align}
where $l_j$ is the mean of the $j$-th constituent consistent Gaussian density and $d_1, d_2, \ldots, d_M$ are real numbers such that $d_1 + d_2 + \ldots + d_n = 1$.
The entropy of the BICM channel will be
\begin{align}
H(c_{BICM(h)}) = \sum_{j = 1}^M d_j \Big[1 - J(\sqrt{2l_j}) \Big].
\label{Eqn_BICM_Entropy}
\end{align}
While for BPSK modulation we have $M=1$, $d_1=1$, and $l_1 = 2/\sigma^2$,
for other modulation schemes, $l_j$ and $d_j$ in \cref{Eqn_GA_mixture} 
are obtained as per the approximations suggested in~\cite{WiMax_standard_Book}.
%
Note that for non-Gray mapping, finding the exact or approximate distribution of $L_i$ becomes difficult.
%

%
\subsection{EBP-GEXIT chart for GLDPC/DGLDPC codes over the BICM channel}
\label{subsection_G_DG_LDPC_EBP}

We first extend the EBP-GEXIT chart proposed for LDPC codes over BAWGN channel to GLDPC and DGLDPC over BICM channel.
Let $f_{C}(.)$ and $f_{V}(.)$ be the functions corresponding to the processing done by a randomly chosen CN and VN respectively while performing the BP decoding.
Then similar to \cref{Eqn_DE_general_LDPC}, the DE equation for BICM$(h)$ is given by $a^{BP,l} = c_{BICM(h)} \star f_V(f_C(a^{BP,l-1}))$,  
where $a^{BP,l}$ is the density of the message passed by a randomly chosen VN to CN.
Note that for an irregular LDPC code, $f_{C}(.) = \rho(.)$ and $f_{V} = \lambda(.)$~\cite[Thm.~4.97]{Urbanke_Richardson_MCT_Book} and 
for GLDPC and DGLDPC codes, $f_{C}(.)$ and $f_{V}(.)$ needs to be obtained numerically~\cite{Liva_Quasi_Cyclic_2008,Yige_DG_ISIT_2006}.

%
To find a FP density pair (see \cref{subsection_EBP_GEXIT_preliminaries}) corresponding to the given BICM channel with the $L$-density $c_{BICM(h)}$, we need to find all possible densities $a$  that satisfy 
%
\begin{align}
a = c_{BICM(h)} \star f_V(f_C(a))).
\label{Eqn_FP_DGLDPC}
\end{align}
%
%
For the given fixed-density pair ($a$, $c_{BICM(h)}$), the EBP-GEXIT function is given by
\begin{align} 
%
%
g^{EBP}(h) \coloneqq \int_{-\infty}^{\infty}  \Lambda(f_C(a))(z) l^{c_{BICM(h)}}(z) \mbox{d}z,	
\label{Eqn_EBP_GEXIT_DGLDPC_BICM}
\end{align}
where $l^{BICM(h)}(z)$ is the GEXIT kernel for BICM($h$). From \cref{Eqn_GA_mixture}, $l^{BICM(h)}(z)$ can be expressed in terms of the GEXIT kernel of binary input AWGN channel
as follows
\begin{align}
l^{c_{BICM(h)}}(z) = \sum_{j=1}^M a_j l^{c_{BAWGN(h_j)}}(z),
\label{Eqn_GEXIT_Kernel_BICM}
\end{align}
where $h_j = 1 - J(\sqrt{2l_j})$ and $l^{c_{BAWGN(h_j)}}(z)$ is defined in \cref{Eqn_BAWGN_kernel}.
%

%
\subsection{Numerical computation of EBP-GEXIT chart for GLDPC/DGLDPC codes}
\label{subsection_G_DG_LDPC_EBP_numerical}

In order to compute the EBP-GEXIT chart in a tractable manner, key steps are to to compute \cref{Eqn_FP_DGLDPC} and \cref{Eqn_EBP_GEXIT_DGLDPC_BICM} in a computationally tractable manner, which are explained next.

\subsubsection{Numerical computation of FP density in \cref{Eqn_FP_DGLDPC}}
\label{subsubsection_complete_FP_family_LDPC}

We assume that the density $a$ in \cref{Eqn_FP_DGLDPC} is consistent normal, i.e., for some real number $m_a$, the density $a$ is $\mathcal{N}(m_a,2m_a)$. 
This consistent Gaussian assumption proposed by Chung et al.~\cite{GA_Urbanke_2001} is also used for classical EXIT charts analysis \cite{Brink_EXIT_2004}.
Using the consistent Gaussian assumption for $a$ is the key idea that simplifies the operations required towards finding the EBP-GEXIT curve.
We next explain how FP density in \cref{Eqn_FP_DGLDPC} can be efficiently approximated using a classical EXIT-like mono-dimensional FP equation.

Similar to the EXIT-chart analysis, we consider the mutual information (MI) $I_{E_v}$ between the LLRs and their corresponding VNs bits. It is given by $I_{E_v} = J(\sqrt{2m_a})$ where $2m_a$ is the variance of the density $a$ and $J(.)$ is defined in \cref{Eqn_BAWGN_Entropy}.
Note that $J(.)$ is a one-to-one function and this implies that the density $a$ can be uniquely determined from it.
Similarly, let $I_{E_c}$ be the MI corresponding to the density $f_C(a)$ of the CN to VN message and suppose $I_{E_c} = \Gamma^C(I_{E_v})$. Note that in notation $\Gamma^C(.)$, we have used the alphabet $\Gamma$ to indicate the transfer function
and the subscript $C$ is used for the CN processing.
The MI corresponding to the VN to CN message 
is a function of $h$ and $I_{E_c}$, denoted by $\Gamma^V(I_{E_c},h)$.
Using this EXIT based mono-dimensional representation, the FP density equations \cref{Eqn_FP_DGLDPC} can be equivalently stated as follows
\begin{align}
I_{E_v} = \Gamma^V \Big( \Gamma^C \big( I_{E_v} \big), h  \Big).
\label{Eqn_FP_GLDPC_mono}
\end{align}
%


%
For the given BICM channel with entropy $h$, the FP density pairs now consists of all those consistent normal densities $a$
such that the corresponding $I_{E_v} = J(\sqrt{2m_a})$ satisfy \cref{Eqn_FP_GLDPC_mono}.
Observe that the FP density in \cref{Eqn_FP_DGLDPC} is now represented by a FP equation \cref{Eqn_FP_GLDPC_mono} since both $I_{E_v}$ and $h$ are scalars. 
All possible pairs $a$ and $c_{BICM(h)}$ that satisfy \cref{Eqn_FP_DGLDPC} can be found efficiently via a grid search by varying $I_{E_v}$ and $h$ in the range $[0,1]$\footnote{One can also use any other efficient methods to find the fixed-points.}. 
This simplifies the process of finding the FP density pairs in \cref{Eqn_FP_DGLDPC}.

For the irregular LDPC codes, the operations $\Gamma^C(.)$ and $\Gamma^V(., .)$ can be simplified as follows
\begin{equation}
\begin{aligned}
\Gamma^C(I_{E_v})  &= \sum_{j} \rho_j \Big( 1 - J\Big[\sqrt{(j-1)[J^{-1}(1-I_{E_v})]^2} \Big] \Big) \\
\Gamma^V(I_{E_c},h) &= \sum_i \lambda_i \sum_{j=1}^M  d_j J \left[ \sqrt{ (i-1) \big[J^{-1}(I_{E_c})\big]^2 + 2l_j }\right],
\end{aligned}
\label{Eqn_DE_irregular_f_C_f_V}
\end{equation}
where $I_{E_c} = \Gamma^C(I_{E_v})$ and $2l_j$ is the variance of the $j$-th constituent density in the mixture $c_{BICM(h)}$.
For GLDPC and DGLDPC codes, $\Gamma^C(.)$ and $\Gamma^V(., .)$ are evaluated point-wise by means of Monte Carlo simulations and stored before computation of \cref{Eqn_FP_GLDPC_mono}~\cite{Liva_Quasi_Cyclic_2008, Yige_DG_ISIT_2006}.
%


%
\subsubsection{Numerical computation of the EBP-GEXIT function
}

For the given BICM$(h)$, let $\mathcal{S}_h$ be the set of all possible $I_{E_v} \in [0,1]$ that satisfy \cref{Eqn_FP_GLDPC_mono}. Recall that corresponding to each $I_{E_v}$ there is a density $a = \mathcal{N}(m_a,2m_a)$ with $I_{E_v} = J(\sqrt{2m_a})$.
Each $a$ corresponding to $I_{E_v} \in \mathcal{S}_h$ provides a point on the EBP-GEXIT curve that is computed using \cref{Eqn_EBP_GEXIT_DGLDPC_BICM}. and \cref{Eqn_GEXIT_Kernel_BICM}. We now provide tractable computation of these equations.
We first explain calculations towards $\Lambda(f_C(a))$ under our Gaussian assumption.
For any $I_{E_v} \in \mathcal{S}_h$, suppose $I_{E_c} = \Gamma^C\big(I_{E_v}\big)$ and $m_b = [J^{-1}(I_{E_c})]^2/2$.
This implies that the density $f_C(a)$ of a randomly chosen message from CNs is consistent Gaussian with mean $m_b$, i.e., $f_C(a) = \mathcal{N}(m_b,2m_b)$.
For a VN of degree $j$, the density obtained by taking the convolution of the input density $j$ times
is also a consistent Gaussian density of mean $jm_b$.
Let us denote this density by $b_j = \mathcal{N}(jm_b,2jm_b)$.
The density $\Lambda(f_C(a))$ is thus the mixture of densities $b_j$ given by $%
\Lambda(b)(f_C(a)) = \sum_j \Lambda_j b_j(z)$.
%
%
Substituting this in \cref{Eqn_EBP_GEXIT_DGLDPC_BICM} we get,
\begin{equation}
\begin{aligned}
g^{EBP}(h) = \int_{-\infty}^{\infty} \Big[ \sum_j \Lambda_j b_j(z) \Big] l^{c_{BICM(h)}}(z) \mbox{d}z
	= \sum_j \Lambda_j  \mathbb{E}_{b_j} \Big[ l^{c_{BICM(h)}}(z) \Big],
%
\end{aligned}
\label{Eqn_B_j_EBP_1}
\end{equation}
where $\mathbb{E}_{b_j}[.]$ is now expectation over the Gaussian density $b_j$.
The expectation $\mathbb{E}_{b_j} [ l^{c_{BICM(h)}}(z) ]$ can be computed efficiently using the Gauss-Hermit quadrature weights as follows \cite{Gauss_Quadrature_book}:
\begin{itemize}
\item Let $H_d$ be the Hermite polynomial of degree $d$ with roots $k_1, k_2, \ldots, k_d$, for some $d \in \mathbb{Z}$.
\item Let $z_i = \sqrt{4jm_b} k_i + jm_b$. Then an approximate value of $\mathbb{E}_{b_j} \big[ l(c_{BICM(h)}(z)) \big]$ is given by
\begin{align}
\mathbb{E}_{b_j} \big[ l(c_{BICM(h)}(z)) \big] \approx
\frac{1}{\sqrt{\pi}} \sum_{i=1}^d  \frac{2^{d-1} d! \sqrt{\pi}}{d^2 [H_{d-1}(k_i)]^2} l^{c_{BICM(h)}}(z_i),
\label{Eqn_Gauss_Hermit_apprx}
\end{align}
where $l^{c_{BICM(h)}}(z_i)$ is defined in \cref{Eqn_GEXIT_Kernel_BICM} and can either be computed using numerical integration or using the approximation suggested in Appendix~C.
%
%
\end{itemize}
%

%
To summarize, the consistent Gaussian assumption enables the computation of the complete FP family (via grid search) and the evaluation of the EBP-GEXIT function (via Gauss-Hermit quadrature weights) computationally feasible.
Detailed steps are provided in \cref{Algorithm_EBP_GEXIT_BICM}.
%
%

\begin{algorithm}
\caption{\label{Algorithm_EBP_GEXIT_BICM} EBP-GEXIT chart for LDPC/GLDPC/DGLDPC codes}
\begin{enumerate}[(1)]
\item {\bf Choose} $h \in [0,1]$ and
let $c_{BICM(h)}$ be the $L$-density corresponding to $\mbox{BICM}(h)$.
\item {\bf Find} $\mathcal{S}_h := \big\{ I_{E_v}: \mbox{s.t.~} I_{E_v} \mbox{ satisfies \cref{Eqn_FP_GLDPC_mono}} \big\}$, via grid search 
of $I_{E_v}$ in the range $[0,1]$. 

\item {\bf Compute} $g^{EBP}(h)$ using \cref{Eqn_B_j_EBP_1}  
for the set of densities $a$ corresponding to each $I_{E_v} \in \mathcal{S}_h$.
%
\item {\bf Plot} all possible values $g^{EBP}(h)$ obtained in step (3) versus the chosen $h$.
\item {\bf Repeat} the process for various values of $h \in [0,1]$.
%
%
\end{enumerate}
\end{algorithm}
%

%
\begin{remark}
On contrary to the definition of complete FP family (see Section~\ref{subsection_EBP_GEXIT_preliminaries}), $a$ and $c_{BICM(h)}$ pairs obtained using Algorithm~\ref{Algorithm_EBP_GEXIT_BICM}
are \textit{not} parameterized by some $x \in [0,1]$, since we find these pairs exhaustively. 
However it can be  easily 
verified that $H(a) = x$ for some $x \in [0,1]$ and the set of $a$ and $c_{BICM(h)}$ obtained
do form a complete FP family.
\hfill $\square$
\end{remark}

%
\subsection{EBP-GEXIT chart for serially concatenated codes over the BICM channel}
\label{subsection_serial_turbo_EBP_Gray}

%
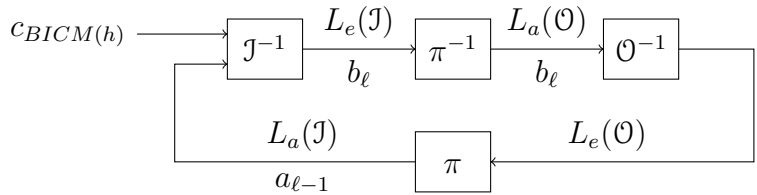
\begin{figure}[t]

\begin{center}

    \begin{tikzpicture}
    
    \draw [->] (-1.2,0.2) -- (0,0.2);  	     	  
    \draw (0,-0.4) rectangle (1,0.4);
    
    \draw [->] (1,0) -- (1+1.5,0);  	     	      
    \draw (1+1.5,-0.4) rectangle (1+1.5+1,0.4);
    
    \draw [->] (1+1.5+1,0) -- (1+1.5+1+1.5,0);  
    \draw (1+1.5+1+1.5,-0.4) rectangle (1+1.5+1+1.5+1,0.4);
    
    \draw [->] (1+1.5+1+1.5+1,0) -- (1+1.5+1+1.5+1+1,0) -- (1+1.5+1+1.5+1+1,-1.5) -- (1+1.5+1,-1.5);  	
    
    \draw (1+1.5,-0.4-1.5) rectangle (1+1.5+1,0.4-1.5);
    
    \draw [->] (1+1.5,-1.5) -- (-0.7,-1.5) -- (-0.7,-0.2) -- (0,-0.2);  	

    \node [above] at (0.5,-0.3) {$\mathcal{I}^{-1}$};
    \node [above] at (0.5+2.5,-0.3) {$\pi^{-1}$}; 	 
    \node [above] at (0.5+2.5+2.5,-0.3) {$\mathcal{O}^{-1}$}; 	 
    \node [above] at (0.5+2.5,-0.3-1.5) {$\pi$};

    \node [left] at (-1.2,0.2) {$c_{BICM(h)}$}; 	 
    
    \node [above] at (1+0.75,0) {$L_e(\mathcal{I})$};
    \node [below] at (1+0.75,0) {$b_{\ell}$}; 	 
	  
    \node [above] at (1+0.75+2.5,0) {$L_a(\mathcal{O})$};
    \node [below] at (1+0.75+2.5,0) {$b_{\ell}$}; 	 	  
	
    \node [above] at (1+0.75+3.3,0-1.5) {$L_e(\mathcal{O})$};

    \node [above] at (1,0-1.5) {$L_a(\mathcal{I})$};
    \node [below] at (1,0-1.5) {$a_{\ell-1}$};

    \end{tikzpicture}
\end{center}    
\vspace{-0.7cm}
\caption{Block diagram of an iterative decoder of a serially concatenated system}
\vspace{-0.7cm}
\label{Figure_serially_concat_turbo}
\end{figure}
%

We first derive an expression for the EBP-GEXIT chart for SC-TC that is inspired from \cite{GEXIT_Turbo_Urbanke_2005}. 
The block diagram of a classical turbo decoder \cite{i2008bit} of a
SC-TC system is shown in \cref{Figure_serially_concat_turbo}.
Observations from the BICM channel with $L$-density $c_{BICM(h)}$ are  given to the inner decoder $\mathcal{I}^{-1}$ as an \emph{a priori} LLRs, denoted by $L_a(\mathcal{I})$. 
The inner decoder performs the BCJR algorithm \cite{Urbanke_Richardson_MCT_Book} and provides the extrinsic LLRs $L_e(\mathcal{I})$ of the inner-decoded bits. 
After interleaving, these constitute the \emph{a priori} LLRs $L_a(\mathcal{O})$ of the outer-coded bits of the decoder $\mathcal{O}^{-1}$. 
The outer decoder also runs the BCJR algorithm and provides the extrinsic LLRs $L_e(\mathcal{O})$ of the outer-coded bits. 

To find the EBP-GEXIT chart, we need to first consider the DE equations for SC-TC (see~\cite[Problem~6.7]{Urbanke_Richardson_MCT_Book}).
Let $b_{\ell}$ and $a_{\ell}$ denote the densities of $L_e(\mathcal{I})$ (or $L_a(\mathcal{O})$) and $L_e(\mathcal{O})$ (or $L_a(\mathcal{I})$) in the ${\ell}$-th iteration of BP decoding.
Note that $b_{\ell}$ is a function of the $L$-density $c_{BICM(h)}$ from the demodulator and the density $a_{\ell-1}$ of \textit{a priori} LLRs available from the outer code in the $\ell-1$-th iteration. Let $b_{\ell} = f_{\mathcal{I}}(c_{BICM(h)}, a_{\ell-1})$, where $f_{\mathcal{I}}(.,.)$ denote the density transfer function corresponding to the processing done by $\mathcal{I}$.
Similarly, let $f_{\mathcal{O}}(.)$ denote the density transfer function corresponding to $\mathcal{O}$, i.e., $a_{\ell} = f_{\mathcal{O}}(b_{\ell})$.
The DE equation for SC-TC described in \Fig~\ref{Figure_serially_concat_turbo} will be $a_{\ell} = f_{\mathcal{O}}\Big(f_{\mathcal{I}}(c_{BICM(h)}, a_{\ell-1})\Big)$\footnote{While DE equation of \cite[Prob.~6.7]{Urbanke_Richardson_MCT_Book} considers separate density transfer functions for systematic and parity bits, we consider the combined transfer function $\Gamma^{\mathcal{O}}(.)$ corresponding to the LLRs of the complete codeword (i.e. both systematic and parity bits).}.
For the given $h$, density $a$ is called as a \textit{FP density} if 
%
\begin{align}
a = f_{\mathcal{O}}\Big(f_{\mathcal{I}}(c_{BICM(h)}, a)\Big).
\label{Eqn_FP_BICM_SCC}
\end{align}
Analogous to parallel concatenation \cite{GEXIT_Turbo_Urbanke_2005}, given a FP density pair $c_{BICM(h)}$ and $a$, the EBP-GEXIT function $g^{BP}(h)$ 
for $\mathcal{S}(\mathcal{O}, \mathcal{I})$ is given by
\begin{align} 
g^{EBP}(h) \coloneqq \int_{-\infty}^{\infty}  f_{\mathcal{I}}(c_{BICM(h)}, a)(z) l^{BICM(h)}(z) \mbox{d}z,	
\label{Eqn_EBP_GEXIT_SCC_BICM}
\end{align}
where recall that $l^{BICM(h)}(z)$ is the GEXIT kernel for BICM($h$) (see \cref{Eqn_GEXIT_Kernel_BICM}). 
%



It is known that the density transfer functions $f_{\mathcal{I}}(.,.)$ and $f_{\mathcal{O}}(.)$ of \cref{Eqn_FP_BICM_SCC} are required to be computed numerically~\cite[Sec.~6.5]{Urbanke_Richardson_MCT_Book} and this may make finding the FP density pairs of and computation of $g^{EBP}(h)$ computationally complex.
To simplify these calculations, we propose to use the EXIT function~\cite{ten2001convergence} corresponding to the processing done by the inner and outer codes. 
Let $I_{A_\mathcal{I}}, I_{E_\mathcal{I}}, I_{A_\mathcal{O}},$ and $I_{E_\mathcal{O}}$ be the MI between the LLRs $L_a(\mathcal{I}), L_e(\mathcal{I}), L_a(\mathcal{O}),$ and $L_e(\mathcal{O})$ and the corresponding bits respectively.
%
%
Suppose $\Gamma^{\mathcal{I}}(.,.)$ and $\Gamma^{\mathcal{O}}(.)$ to denote the MI EXIT function for $\mathcal{I}$ and $\mathcal{O}$ respectively, with $I_{E_\mathcal{I}} = \Gamma^{\mathcal{I}}(h,I_{A_\mathcal{I}})$ and $I_{E_\mathcal{O}} = \Gamma^{\mathcal{O}}(I_{A_\mathcal{O}})$.
Using this, the FP equation corresponding to \cref{Eqn_FP_BICM_SCC} will be $I_{A_\mathcal{I}} = \Gamma^{\mathcal{O}}\big(\Gamma^{\mathcal{I}}\big(h, I_{A_\mathcal{I}}\big)\big)$.
%
Similar to the previous section, FP density of \cref{Eqn_FP_BICM_SCC} can be obtained by finding the fixed-points of $I_{A_\mathcal{I}} = \Gamma^{\mathcal{O}}\big(\Gamma^{\mathcal{I}}\big(h, I_{A_\mathcal{I}}\big)\big).$ since $I_{A_\mathcal{I}} \in [0,1]$ and $h \in [0,1]$.
To find the EBP-GEXIT function corresponding to $h$ and $I_{A_\mathcal{I}}$, we project $I_{A_\mathcal{I}}$ on $\mathcal{N}(m_a,2m_a)$ with $m_a = J^{-1}(I_{A_\mathcal{I}})^2/2$ and the integration in \cref{Eqn_EBP_GEXIT_SCC_BICM} can be computed efficiently using Gauss-Hermit quadrature weights~\cite{Gauss_Quadrature_book}. 

%
\section{EBP-GEXIT chart over the AWGN channel with non-Gray mapping}
\label{Section_EBP_non_Gray}

In this section, we consider the case when modulated symbols are mapped according to any non-Gray mapping. Note that for any non-Gray mapping, the EXIT chart of the detector is not flat and hence the decoding of the system illustrated in \Fig~\ref{Figure_general_digi_comm_system} becomes doubly iterative~\cite{ten2003design}. This implies that for LDPC codes one needs to iterate between VNs and CNs but also between the detector and the decoder (similarly for SC-TC). 
Hence for the computation of the EBP-GEXIT chart we need to consider the complex input AWGN channel. 
We first extend the existing results to obtain an expression for the GEXIT function for non-binary complex input AWGN channel and then provide a method for its tractable computation.

\subsection{GEXIT function for non-binary complex-input AWGN channel}
\label{subsection_GEXIT_non_Gray}

We make use of the definition of the GEXIT function defined in \cite{GAT_Urbanke_2009} and follow the approach proposed in \cite[Sec.~III]{Pfister_ISI_2012} to derive an expression for the GEXIT function for non-binary complex-input AWGN channel. 
We first introduce some notation that we shall need in this section.
Corresponding to the $t$-th transmitted symbol $x_t \in \mathbb{X}$, define a vector $\phi_t$ of length $|\mathbb{X}|$ as follows
\begin{align}
\phi_t \coloneqq \Big[ \mathbb{P}(X_t = \xi_1|\mathbf{y}_{\sim t}) \mbox{~~}  \mathbb{P}(X_t = \xi_2|\mathbf{y}_{\sim t}) \mbox{~~} \ldots \mbox{~~} \mathbb{P}(X_t = \xi_{|\mathbb{X}|}|\mathbf{y}_{\sim t}) \Big]
\label{Eqn_Phi_t_definition}
\end{align}
where $\mathbf{y}_{\sim t} = [ y_1 \mbox{~} \ldots \mbox{~} y_{t-1} \mbox{~} y_{t+1} \mbox{~} \ldots \mbox{~} y_N ]$ and $t = 1, 2, \ldots, N$.
Observe that $\phi_t$ corresponds to the likelihood of $X_t$ given all received symbols except the $t$-th symbol. 
In the presence of an ideal interleaver between the channel code and the modulator (see \Fig~\ref{Figure_general_digi_comm_system}), 
$X_1, X_2, \ldots, X_N$ can be assumed to be independent and hence $\mathbb{P}(X_t = \xi_j|\mathbf{y}_{\sim t})$ will not depend on $\mathbf{y}_{\sim t}$ for any $1 \leq j \leq |\mathbb{X}|$. This implies that $\mathbb{P}(X_t = \xi_j|\mathbf{y}_{\sim t})$ will be a function of the \textit{a priori} knowledge available about $X_t$. We obtain a general expression for the GEXIT function and make use of this independent assumption for its numerical computation. 
Let $\Phi_t$ denotes the random vector corresponding to $\phi_t$.
Let $f_{t,\xi}$ be the distribution of vector $\Phi_t$ under the condition $X_t = \xi$, i.e., $f_{t,\xi}(\phi_t) \coloneqq \mathbb{P}\big[ \Phi_t = \phi_t|X_t = \xi  \big]$, where $\xi \in \mathbb{X}$.
Let $\phi_{t,[\xi]}$ denotes the entry in vector $\phi_t$ that corresponds to likelihood of the symbol $\xi \in \mathbb{X}$, i.e., $\phi_{t,[\xi]} \coloneqq \mathbb{P}(X_t = \xi|\mathbf{y}_{\sim t})$.
The notation $[\xi]$ in $\phi_{t,[\xi]}$ denotes the index of the entry in the vector $\phi_t$ associated with $\xi \in \mathbb{X}$.
We now derive an expression for $|\mathbb{X}|$-ary complex-input memoryless AWGN channel. 

\begin{theorem}
\label{Theorem_GEXIT_nonbin_complex_AWGN_our}
Consider $\phi_t, f_{t,\xi}(\phi),$ and $\phi_{t,[\xi]}$ as defined above (see \cref{Eqn_Phi_t_definition}).
%
%
Then the GEXIT function $g(h)$ for $|\mathbb{X}|$-ary complex-input memoryless AWGN channel with entropy $h$ is given by $g(h) = \frac{1}{N} \sum_{t=1}^N \frac{A_t(h)}{B_t(h)}$, where $A_t(h)$ and $B_t(h)$ are given by
\begin{align*}
A_t(h) &= \sum_{\xi \in \mathbb{X}}  \int_{\phi_t}f_{t,\xi}(\phi_t)
%
\int_{y_t} 
\frac{e^{-\frac{|y_t-\xi|^2}{2 \sigma^2}} }{2\pi\sigma^2}
\left[ |y_t-\xi|^2 - 2\sigma^2\right] \\
& \mbox{~~~~~~~~~~~~~~~~~~~~~~~}\log_2 
\Bigg\{
\sum_{\xi^{\prime} \in \mathbb{X}} 
\frac{\phi_{t,[\xi^{\prime}]}}{\phi_{t,[\xi]}} 
\exp \left[ \frac{|y_t-\xi|^2 - |y_t-\xi^{\prime}|^2}{2 \sigma^2}  \right]
\Bigg\} 
\mathrm{d}y_t
%
\mathrm{d}\phi \\
B_t(h) &= \sum_{\xi \in \mathbb{X}}  
%
\int_{y_t} 
\frac{e^{-\frac{|y_t-\xi|^2}{2 \sigma^2}} }{2\pi\sigma^2}
\left[ |y_t-\xi|^2 - 2\sigma^2\right] 
\log_2 
\Bigg\{
\sum_{\xi^{\prime} \in \mathbb{X}} \exp \left[ \frac{|y_t-\xi|^2 - |y_t-\xi^{\prime}|^2}{2 \sigma^2}  \right]
\Bigg\} 
\mathrm{d}y_t.
\end{align*}
\end{theorem}
The proof is given in Appendix~A. 
The fraction $g_t(h) \coloneqq A_t(h)/B_t(h)$ in \cref{Theorem_GEXIT_nonbin_complex_AWGN_our} is termed as the $t$-th GEXIT function~\cite{Pfister_ISI_2012}.
%
%
We next find the EBP-GEXIT function. Similar to \cref{Eqn_Phi_t_definition}, consider 
$\Phi_t^{BP,l}$ corresponding to the likelihood of $X_t$ given $\mathbf{y}_{\sim t}$ in the $l$-th round of BP decoding and let $f_{t,\xi}^{BP,l}$ be the density of $\Phi_t^{BP,l}$ under the condition $X_t = \xi$. The BP-GEXIT function $g^{BP,l}(h)$ in the $l$-th round of BP-decoding is  obtained by substituting $f_{t,\xi} = f_{t,\xi}^{BP,l}$ in \cref{Theorem_GEXIT_nonbin_complex_AWGN_our} and the BP-GEXIT function is defined as $g^{BP}(h) \coloneqq \lim_{l \rightarrow \infty }g^{BP,l}(h)$~\cite{GAT_Urbanke_2009}.
The EBP-GEXIT function is obtained by computing $g^{BP}(h)$ for each FP density pair.
%
To apply Maxwell construction, we next provide the area theorem for $|\mathbb{X}|$-ary complex-input AWGN channel.

\begin{theorem}
\label{Theorem_area_theorem_nonbin_AWGN_our}
Consider a family of channel codes of rate $k/n$ and transmission using the digital communication system of \Fig~\ref{Figure_general_digi_comm_system}
over $\{\mbox{AWGN}(h)\}_h$ using $2^m$-ary modulation scheme.
%
Then 
\begin{align*}
\frac{1}{m} \int_{0}^{m} g(h) \mathrm{d}h = \frac{k}{n}.
\end{align*}
\end{theorem}
The proof is given in Appendix~B.
Having obtained an expression for the GEXIT function, we next consider a tractable computation of the EBP-GEXIT chart. 
 
\subsection{Numerical computation of the complete FP family}
%
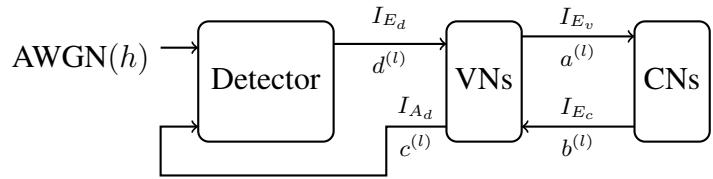
\begin{figure}[t]

\begin{center}

    \begin{tikzpicture}
    
      \draw [rounded corners, thick] (0,-0.8) rectangle (1.8,0.8);	          	    \node [above] at (0.9,-0.25) {Detector}; 	   	   	   	   	  
    
      \draw [->,thick] (-0.5,0.45) -- (0,0.45);	  
      \node [left] at (-0.5,0.3) {AWGN$(h)$};

      \draw [->,thick] (1.8,0.5) -- (1.8+1.5,0.5);	  
      \draw [rounded corners, thick] (1.8+1.5,-0.8) rectangle (1.8+1.5+1,0.8);	              \node [above] at (0.5+1.8+1.5,-0.25) {VNs}; 	   	   	   	   	  
      
      \node [above] at (1.8+1.5+1+0.75,0.55) {\footnotesize$I_{E_{v}}$}; 	    
      \node [below] at (1.8+1.5+1+0.75,0.65) {\footnotesize$a^{(l)}$}; 	    	
      \node [above] at (1.8+1.5+1+0.75,-0.65) {\footnotesize$I_{E_{c}}$}; 	    
      \node [below] at (1.8+1.5+1+0.75,-0.55) {\footnotesize$b^{(l)}$}; 	    	

      \node [above] at (1.8+0.75,0.55) {\footnotesize$I_{E_{d}}$}; 	    
      \node [below] at (1.8+0.75,0.55) {\footnotesize$d^{(l)}$}; 	    	      
    
      \node [above] at (1.8+1.1,-0.65) {\footnotesize$I_{A_{d}}$}; 	    
      \node [below] at (1.8+1.1,-0.55) {\footnotesize$c^{(l)}$}; 	    
    
      \draw [->,thick] (1.8+1.5+1,0.6) -- (1.8+1.5+1+1.5,0.6);	  
      \draw [rounded corners, thick] (1.8+1.5+1+1.5,-0.8) rectangle (1.8+1.5+1+1.5+1,0.8);	
      \node [above] at (0.5+1.8+1.5+1+1.5,-0.25) {CNs}; 	   	   	   	   	  
      
      \draw [<-,thick] (1.8+1.5+1,-0.6) -- (1.8+1.5+1+1.5,-0.6);	  
      
      
      \draw [->,thick] (1.8+1.5,-0.6) -- (2.5,-0.6) -- (2.5,-0.6-0.65) -- (-0.5,-0.6-0.65) -- (-0.5,-0.6) -- (0,-0.6);

    \end{tikzpicture}

\end{center}    

\caption{Considered decoding scheduling for the LDPC-coded serially concatenated scheme}
\label{Figure_AWGN_PSK_ASK_Arbitrary_mapping}  	  
\vspace{-0.4in}
\end{figure}
%

We now find the complete FP family for LDPC codes. Towards this, we 
need to consider the DE equations of the system illustrated in \cref{Figure_AWGN_PSK_ASK_Arbitrary_mapping}.
In one iteration of decoding, first a message is passed from the detector to VNs. This message is a function of the channel parameter $h$ and the incoming message from VNs in the previous iteration. VNs then pass messages to CNs, which are then passed back to VNs after CN processing. Finally, an average message from the VNs is sent back to the detector.
Let $a^{(l)}, b^{(l)}, c^{(l)},$ and $d^{(l)}$ be the density of a randomly chosen message from VNs-to-CNs, CNs-to-VNs, VNs-to-detector, and detector-to-VNs in the $\ell$th iteration (note that this scheduling is in spirit equivalent to the combined VN and detector approach proposed in~\cite[Fig.~5]{Brink_Modulation_2004}).
%
The DE equations are then given by
\begin{equation}
\begin{aligned}
%
a^{(\ell)} &= d^{(\ell-1)} \star f_V(b^{(\ell-1)}), \mbox{~~~} b^{(\ell)} = f_C(a^{(\ell)})  \\
c^{(\ell)} &= f_V^{\prime}(b^{(\ell)}), \mbox{~~~~~~~~~~~~~~~}d^{(\ell)} = f_D(c^{(\ell)}, h),
\end{aligned}
\label{Eqn_DE_Serially_concatenated_True}
\end{equation}
where the function $f_D(.,.)$ depends on the underlying detector, $f_V^{\prime}(.)$ correspond to the VN-to-detector processing and recall that $f_V(.)$ and $f_C(.)$ correspond to VN and CN processing respectively.
Note that the function $f_D(.,.)$ depends on the channel parameter $h$.
In (\ref{Eqn_DE_Serially_concatenated_True}) when $a^{(\ell)} = a^{(\ell-1)}$ then such a density will be a fixed point density (see \cref{subsection_EBP_GEXIT_preliminaries}), i.e.,
for the given $h$ the density $a$ is called as fixed point density if it satisfies the following equation
\begin{align}
a = f_D\Big( f_V^{\prime}\big( f_C(a) \big), h \Big) \star f_V \Big( f_C(a) \Big).
\label{Eqn_FP_Serially_concat_True}
\end{align}
%

Similar to \cref{subsubsection_complete_FP_family_LDPC}, we project the densities
in (\ref{Eqn_DE_Serially_concatenated_True}) on their respective MIs.
%
Let $I_{E_v}, I_{E_c}, I_{A_d},$ and $I_{E_d}$ be the MIs corresponding to $a^{(l)}, b^{(l)}, c^{(l)},$ and $d^{(l)}$ respectively.
Similarly consider the respective MI transfer functions denoted by $\Gamma^V(.), \Gamma^C(.), \Gamma^{{\prime}V}(.), $ and $\Gamma^D(.)$.
Using this the DE equations in (\ref{Eqn_DE_Serially_concatenated_True}) can be represented as
\begin{equation}
\begin{aligned}
I_{E_v}^{(\ell)} &= \Gamma^V\big( I_{E_c}^{(\ell-1)}, I_{E_d}^{(\ell-1)}\big), \mbox{~~~~~~~} I_{E_c}^{(\ell)} = \Gamma^C(I_{E_v}^{(\ell)})  \\
I_{A_d}^{(\ell)} &= \Gamma^{\prime V}(I_{E_c}^{(\ell)}), \mbox{~~~~~~~~~~~~~~~~~}I_{E_d}^{(\ell)} = \Gamma^{D}(I_{A_d}^{(\ell)}, h),
\end{aligned}
\label{Eqn_DE_Serially_concatenated_MI}
\end{equation}
%
%
%
%
%
%

From \cref{Eqn_DE_Serially_concatenated_MI}, the FP density in \cref{Eqn_FP_Serially_concat_True} can be expressed as
%
\begin{align}
I_{E_v} = \Gamma^V \bigg( \Gamma^C(I_{E_v}) , \Gamma^{D} \Big( \Gamma^{\prime V}(\Gamma^C(I_{E_v})), h \Big) \bigg).
\label{Eqn_FP_Serially_concat_MI}
\end{align}

Similar to \cref{subsubsection_complete_FP_family_LDPC}, all possible pairs $a$ and $h$ that satisfy FP density in \cref{Eqn_FP_Serially_concat_True} can be found efficiently from \cref{Eqn_FP_Serially_concat_MI} via grid search by varying $I_{E_v}$ and $h$ in the ranges $[0,1]$ and $[0, H(X)]$ respectively (recall that $H(X)$ is the entropy of the input alphabet set $\mathcal{X}$ to the AWGN channel). 
For SC-TC, DE equations similar to \cref{Eqn_DE_Serially_concatenated_True} and \cref{Eqn_DE_Serially_concatenated_MI} can be written and the FP family can be obtained in a similar manner. We skip these details.

\subsection{Numerical computation of the EBP-GEXIT function}
\label{subsection_numerical_computation_EBP_non_Gray}

We now provide a method for numerical computation of the GEXIT function derived in \cref{Theorem_GEXIT_nonbin_complex_AWGN_our}.
In this theorem, as the expression for $A_t(h)$ and $B_t(h)$ is the same for $t = 1,2,\ldots, N$, we get $A_1(h) = \ldots = A_N(h) = A(h)$ and $B_1(h) = \ldots = B_N(h)= B(h)$ and hence
\begin{align}
g(h) = \frac{1}{N}\sum_{t=1}^N \frac{A_t(h)}{B_t(h)} = \frac{A(h)}{B(h)}.
\end{align}

For the sake of convenience, we shall now drop the suffix $t$ from the expression of $A_t(h)$ and $B_t(h)$.
%
We remove the suffix $t$ from the terms $f_{t,\xi}(\phi_t)$, $\Phi_t$, and $Y_t$ as well.
Let us denote the term inside the integration with respect to $y$ in the expression of $A(h)$ by $R_1(y, \phi, \xi, \sigma)$, i.e., 
\begin{equation}
\begin{aligned}
R_1(y, \phi, \xi, \sigma) \coloneqq 
\left[ |y-\xi|^2 - 2\sigma^2\right] 
\log_2 
\Bigg\{
\sum_{\xi^{\prime} \in \mathbb{X}} \frac{\phi_{[\xi^{\prime}]}}{\phi_{[\xi]}} \exp \left[ \frac{|y-\xi|^2 - |y-\xi^{\prime}|^2}{2 \sigma^2}  \right]
\Bigg\}.
%
%
\end{aligned}
\label{Eqn_define_R_func_GEXIT_nonGray_At_Bt}
\end{equation}
Using this $A(h)$ can be written as 
%
\begin{align}
A(h) &= \sum_{\xi \in \mathbb{X}}  \int_{\phi}f_{\xi}(\phi)
\int_{y} \frac{e^{-\frac{|y-\xi|^2}{2 \sigma^2}} }{2\pi\sigma^2}
R_1(y,\phi, \xi, \sigma)
\mathrm{d}y
\mathrm{d}\phi, \nonumber \\
%
&\stackrel{(a)}{=} \sum_{\xi \in \mathbb{X}} \int_{\phi}f_{\xi}(\phi)
\bigg(
\mathbb{E}_{Y|X=\xi} \big[ R_1(Y,\phi, \xi, \sigma) \big]
\bigg)
\mathrm{d}\phi
%
%
\stackrel{(b)}{=} \sum_{\xi \in \mathbb{X}} \mathbb{E}_{\Phi|X=\xi} \Big[ R_2(\Phi, \xi, \sigma) \Big],
\label{Eqn_expectation_phi_condi_Xt}
\end{align}
where $R_2(\Phi, \xi, \sigma) \coloneqq \mathbb{E}_{Y|X=\xi} \big[ R_1(Y,\phi, \xi, \sigma) \big]$.
The equality in $(a)$ is obtained since the integration with respect to $y$ is equal to the expectation of $R_1(y,\phi, \xi, \sigma)$ with respect to the random variable $Y$ under the condition $X=\xi$. 
The equality in $(b)$ is obtained since the integration with respect to $\phi$ is equal to the expectation of $R_2(\phi, \xi, \sigma)$ with respect to the random vector $\Phi$ under the condition $X=\xi$.
For computing $A(h)$, the key step now is to compute $R_2(\phi,\xi,\sigma)$ and its expectation with respect to random variable $\Phi|X=\xi$.
%
These computations are described next.
\begin{itemize}
\item Computing $R_2(\phi,\xi,\sigma)$ defined in \cref{Eqn_expectation_phi_condi_Xt}:
The function $R_2(\phi,\xi,\sigma)$ is given by
\begin{align}
R_2(\phi,\xi,\sigma) = \int_{y} \frac{e^{-\frac{|y-\xi|^2}{2 \sigma^2}} }{2\pi\sigma^2}
R_1(y,\phi, \xi, \sigma)
\mathrm{d}y
\end{align}
Observe that the distribution of complex random variable $Y$ under the condition $X=\xi$ is bivariate Gaussian with mean $\xi$ and variance $\sigma^2$.
This expectation can be computed efficiently via two-dimensional Gauss-Hermit quadrature weights as follows \cite{Gauss_quadrature_multivariate_Jackel}:

\begin{itemize}
\item Suppose $\xi = \xi^r + i \xi^i$ where $\xi^r$ and $\xi^i$ are the real and imaginary parts of $\xi$.
\item Let $H_d$ be the Hermite polynomial of degree $d$ with roots $k_1, k_2, \ldots, k_d$.
\item Let $z(j_1, j_2) = \big[ \sqrt{2} \sigma k_{j_1} + \xi^r \big] + i\big[\sqrt{2} \sigma k_{j_2} + \xi^i \big]$, $w_{j_1} = 2^{d-1} d! \sqrt{\pi}/d^2 [H_{d-1}(k_{j_1})]^2$, and $w_{j_2} = 2^{d-1} d! \sqrt{\pi}/d^2 [H_{d-1}(k_{j_2})]^2$ for $j_1, j_2 = 1,2, \ldots, d$.
Then $R_2(\phi, \xi, \sigma)$ can be approximated as
\begin{align*}
R_2(\phi, \xi, \sigma) \approx
\frac{1}{\pi} 
\sum_{j_1=1}^d
\sum_{j_2=1}^d  
w_{j_1} w_{j_2} R_1\Big[z(j_1, j_2), \phi, \xi, \sigma \Big],
%
%
\end{align*}
where $R_1\big(z(j_1, j_2), \phi, \xi, \sigma \big)$ is defined in \cref{Eqn_define_R_func_GEXIT_nonGray_At_Bt}.
\end{itemize}
%
\item Computing $A(h)$:
%
%
In presence of an ideal interleaver between codewords and the modulator, the set of random variables $X_1, X_2, \ldots, X_N$ can be assumed to be independent. This implies $\Phi_1, \Phi_2, \ldots, \Phi_N$ are also independent.
Without loss of generality we next provide computation steps for any $i$-th vector $\Phi_i$. 
For the system of \Fig~\ref{Figure_general_digi_comm_system}, note that 
$\phi_i$
does not depend on $\mathbf{y}_{\sim i}$ and hence  \cref{Eqn_Phi_definition_modulator} can be simplified to 
\begin{align}
%
%
\phi_i = \Big[ \mathbb{P}(X_i = \xi_1) \mbox{~~} \mathbb{P}(X_i = \xi_2) \mbox{~~} \ldots \mbox{~~} \mathbb{P}(X_i =\xi_{|\mathbb{Z}|}) \Big].
\label{Eqn_Phi_definition_modulator}
\end{align}
Recall $\mathcal{M}$ is the map corresponding to the given $2^m$-ary modulation scheme, i.e., $\mathcal{M}:\mathbf{c}^{\prime}(i) \rightarrow x_i$, where $x_i \in \mathbb{X}$ is a complex constellation symbol 
corresponding to the given modulation scheme 
and 
$\mathbf{c}^{\prime}(i) = [ c_{i,1}^{\prime} \mbox{~} c_{i,2}^{\prime} \mbox{~} \ldots \mbox{~} c_{i,m}^{\prime} ]$ for $i = 1,2, \ldots, N$ (see \cref{Section_System_Model}).
Let $\mathbf{L}_a^{\mathcal{M}}(i) = [ L^{\mathcal{M}}_a(i,1) \mbox{~} L^{\mathcal{M}}_a(i,2) \mbox{~} \ldots \mbox{~} L^{\mathcal{M}}_a(i,m) ]$ be the \textit{a priori} LLRs available at the input of the detector. 
These LLRs are obtained after deinterleaving the VN-to-detector processing.
Since $\mathbf{c}(i) = [ c_{i,1} \mbox{~} c_{i,2} \mbox{~} \ldots \mbox{~} c_{i,m} ]$ denotes the deinterleaved codebit sequence, 
$L^{\mathcal{M}}_a(i,j)$ is given by 
\begin{align}
L^{\mathcal{M}}_a(i,j) = \log \frac{\mathbb{P}(C_{i,j}=0)}{\mathbb{P}(C_{i,j}=1)}, \forall j \in \{ 1, \ldots m\}
\label{Eqn_LLR_a_definition}
\end{align}
%
For each symbol $\xi_l \in \mathbb{X} = \{ \xi_1, \xi_2, \ldots, \xi_{|\mathbb{X}|}\}$, suppose $\mathcal{M}^{-1}(\xi_l) = [ b_{l,1} \mbox{~} b_{l,2} \mbox{~} \ldots b_{l,m} ]$ for $l = 1, 2, \ldots, |\mathbb{X}|$. 
Then the $l$-th entry in $\phi_i$ in \cref{Eqn_Phi_definition_modulator} can be calculated as follows  
\begin{align}
\mathbb{P}(X_i =\xi_l) =\prod_{j = 1}^m \mathbb{P}[C_{i,j} = b_{l,j}],
\label{Eqn_Phi_calculation_for_EBP_nongray}
\end{align}
where $\mathbb{P}[C_{i,j} = b_{l,j}]$ is obtained from \cref{Eqn_LLR_a_definition}.
To compute $A(h) = \sum_{\xi \in \mathbb{X}} \mathbb{E}_{\Phi|X=\xi} \big[ R_2(\Phi, \xi, \sigma) \big]$, we need to find the distribution $f_{\xi}(\phi)$ of $\Phi|X=\xi$ (see \cref{Eqn_expectation_phi_condi_Xt}). 
However finding this multivariate distribution, in general, is not straightforward. Hence we choose to obtain this expectation numerically as follows:
\begin{itemize}
\item Let $I_{A_d}$ be the MI available at the input of the detector (see \cref{Figure_serially_concat_turbo}). 
Project $I_{A_d} \in [0,1]$ on the consistent Gaussian density $\mathcal{N}(m_d,2m_d)$, where $m_d = J^{-1}(I_{A_d})^2/2$. 
\item Generate a sequence of modulated symbols $x_1, x_2, \ldots, x_N$ according to uniform distribution for large enough $N$. (We choose $N = 10000$ in our simulations.)
\item Generate a sequence of \textit{a priori} LLRs $\mathbf{L}_a^{\mathcal{M}}(1), \mathbf{L}_a^{\mathcal{M}}(2), \ldots, \mathbf{L}_a^{\mathcal{M}}(N)$ corresponding to $x_1, x_2, \ldots, x_N$, where each entry in $\mathbf{L}_a^{\mathcal{M}}(i) = \big[ L^{\mathcal{M}}_a(i,1) \mbox{~} L^{\mathcal{M}}_a(i,2) \mbox{~} \ldots \mbox{~} L^{\mathcal{M}}_a(i,m) \big]$ is chosen i.i.d.~according to $\mathcal{N}(m_d,2m_d)$ distribution.
\item Compute the sequence of vectors $\phi_1, \phi_2, \ldots, \phi_N$, where each entry in $\phi_i$ is calculated from $\mathbf{L}_a^{\mathcal{M}}(i)$ using \cref{Eqn_Phi_definition_modulator} and \cref{Eqn_Phi_calculation_for_EBP_nongray}, for $i = 1, 2, \ldots, N$.
\item Given a particular symbol $\xi \in \mathbb{X}$, let $\mathcal{S}_{\xi}$ denotes the set of vectors $\phi_i$ such that the corresponding generated $x_i = \xi$ for $i = 1, 2, \ldots, N$.
\item By approximating expectation of $R_2(\Phi, \xi, \sigma)$ by computing the average over $\mathcal{S}_{\xi}$, the expression of 
$A(h) = \sum_{\xi \in \mathbb{X}} \mathbb{E}_{\Phi|X=\xi} \big[ R_2(\Phi, \xi, \sigma) \big]$ is now approximated by 
%
%
\begin{align*}
%
%
A(h) \approx \sum_{\xi \in \mathbb{X}} \frac{1}{|\mathcal{S}_{\xi}|} \sum_{\phi_i \in \mathcal{S}_{\xi}}R_2\big(\phi_i, \xi, \sigma \big),
\end{align*}
where $R_2\big(\phi_i, \xi, \sigma \big)$ is computed using step 1) as explained above.
\end{itemize} 
%
\item Computing $B(h)$:
When $\phi_{[\xi]} = \phi_{[\xi^{\prime}]}$ for any $\xi, \xi^{\prime} \in \mathbb{X}$, $B(h)$ of \cref{Theorem_GEXIT_nonbin_complex_AWGN_our} can be written 
in terms of $R_2(\phi,\xi,\sigma)$ and computed as $B(h) = \sum_{\xi \in \mathbb{X}} R_2\left(\phi = \left[\frac{1}{|\mathbb{X}|} \mbox{~} \frac{1}{|\mathbb{X}|} \mbox{~} \ldots \mbox{~}\frac{1}{|\mathbb{X}|}\right], \xi, \sigma \right)$.
%
%
%
%
\end{itemize}

%
\section{Spatial coupling analysis of SC-TC}
\label{Section_Spatial_Coupling}

In this section, we provide a procedure for spatial coupling of SC-TC to evaluate their BP thresholds. This formalism was first introduced in \cite{Tarik_Globecomm_2017} and is in spirit analogous to \cite{costello2016new,moloudi2016spatially,moloudi2014spatially}. Inspired from the formalism of spatially-coupled protograph-based LDPC codes, the proposed procedure provides a similar formalism allowing to evaluate spatial coupling parameters such as termination, mapping, choice of base matrices, BP decoding, rate loss, and \emph{wave} effect of LLRs.

As for spatially-coupled protograph-based LDPC codes, spatially-coupled SC-TC can be obtained by the edge spreading rule on the factor-like graph \cite{Tarik_Globecomm_2017}: (i) this latter is duplicated say $L$ times; (ii) then the outer-code encoded bits $\bm{v}$ are divided into $m_s+1$ clusters; (iii) these latter are exchanged between the graph copies by interchanging the ends of homologous sockets following the matrix $B \coloneqq [b_0, b_1, \ldots b_{m_s}] \in [0,1]^{m_s+1}$, where $b_i$ represents the $\bm{v}$ bits fraction passed from the graph copy $t$ to $(t+i)$.

\begin{figure*}[!t]
\captionsetup{justification=centering}
\centering	
	\hspace{-1cm}\begin{subfigure}[c]{0.6\linewidth}
	    \centering
    	\includegraphics[width=1\linewidth]{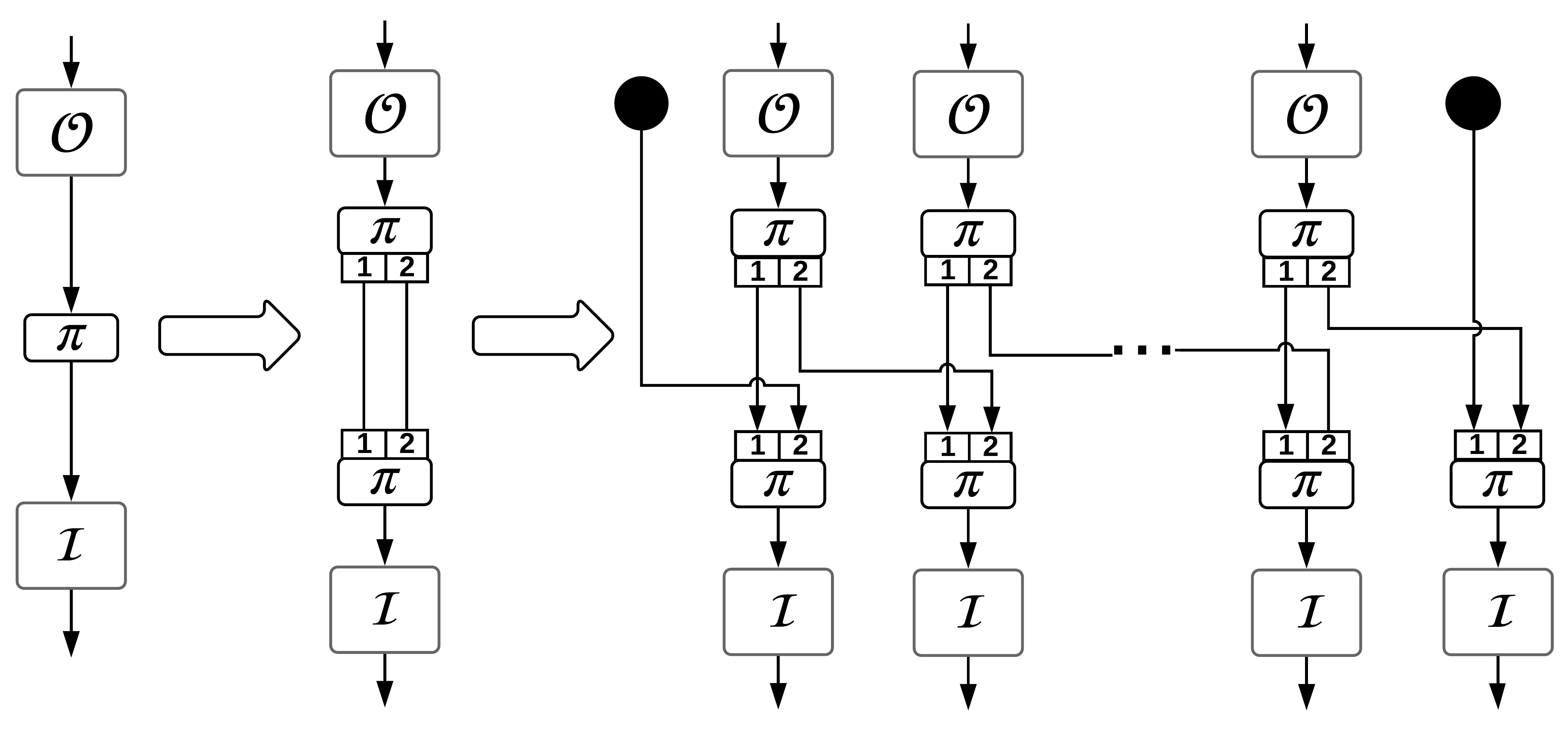}
        \caption{Terminated encoder. Here $B=[0.5, 0.5]$.}
        \label{figSCTCTx}
	\end{subfigure}
	\begin{subfigure}[c]{0.4\linewidth}
	    \centering
    	\includegraphics[width=1\linewidth]{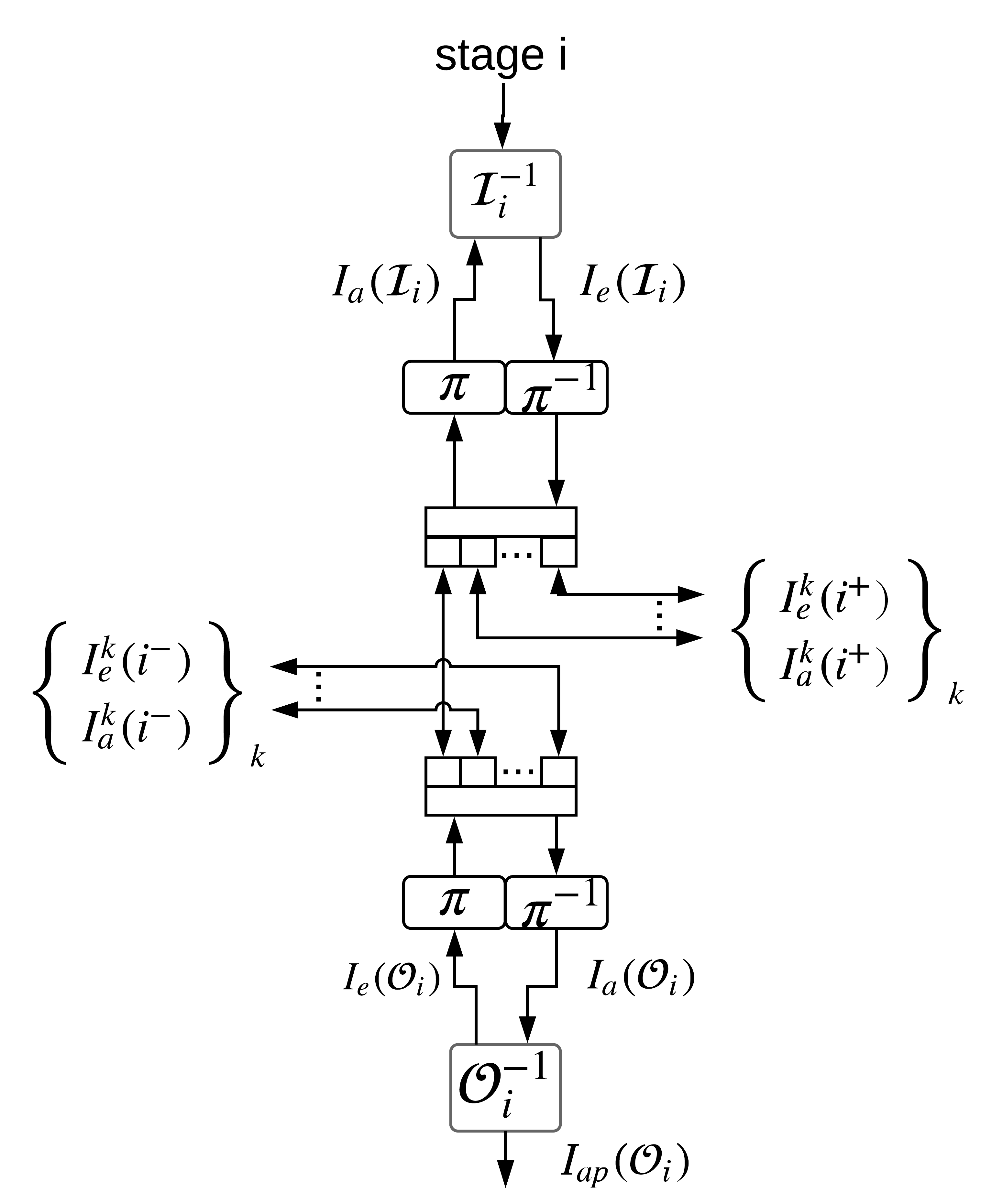}
        \caption{{An arbitrary stage at the receiver}}
        \label{figSCRx}
	\end{subfigure}\hspace{-1.9cm}
\caption{Terminated SC-TC transmitter and receiver}
\label{fig16QAMIter}
\vspace{-0.8cm}
\end{figure*}

Similar to spatially-coupled LDPC codes, $m_s$ is called the syndrome former memory, $L$ the coupling length, and the coupling matrix $B$ verifies the constraint $\sum_{i=0}^{m_s} b_i = 1$. For a better illustration of the described edge spreading rule construction, an example is illustrated in \cref{figSCTCTx}. As one can observe, some clusters will remain unconnected on the rightmost copies and some vacant points on the leftmost copies. Thus suggest that some form of termination should be envisioned. Filling the remaining bundles connections at the boundaries of the obtained graph is classically solved as follows:
\begin{itemize}
\item $m_s$ add inner codes constituents at the rightmost end in order to connect the last remaining bundles.
\item \emph{"padd"} with known information bits at the $m_s$ first and the $m_s$ last stages in order to fill the vacant bundles connection points. These are showed in \cref{figSCTCTx} with black circles.
\end{itemize}

The rate of the coupled scheme is given by $R_L = R - \frac{m_s}{L+m_s}R$. Because of the chosen termination strategy, we induce a rate loss, \emph{i.e.} $\frac{m_s}{L+m_s}R$, that vanishes to $0$ as $L \rightarrow +\infty$. Note that termination methods such as tail-biting \cite{cammerer2016triggering} and code modification \cite{tazoe2012efficient} can also be considered. 
%

\subsection{EXIT analysis of spatially-coupled SC-TC}

Input-output transfer functions of the SISO components can be computed.  These functions represent the MI associated with extrinsic LLR messages at the output of SISO components versus the MI associated with the \emph{a priori} LLR messages.
In cases where DE can be computationally complex or unfeasible, the EXIT chart is a powerful tool to study the asymptotic convergence of concatenated systems under iterative decoding. First introduced in \cite{ten2001convergence}, its idea relies on the fact that the density of  LLRs exchanged during the iterative decoding can be accurately modeled as consistent Gaussian. One can thus evaluate the convergence of the overall system by tracking either the mean or the variance of the LLRs.
In our case, it is not possible to provide analytical expressions of the exchanged LLRs densities for general constituent convolutional codes. Therefore, we propose express EXIT decoding transfer functions of the proposed spatially-coupled SC-TC system under BP decoding (for details refer \cite{Tarik_Globecomm_2017}).

In order to define the main notation, let us consider the $i$-th stage of the spatially-coupled factor-like graph of \cref{figSCRx}. The corresponding notations are defined as follows:
\begin{itemize}
\item all variables corresponding to the stage $i$ are referred to with the subscript $i$;
\item $I_{a}^k(i^+)$ (resp. $I_{e}^k(i^+)$) is the \emph{a priori} (resp. extrinsic) MI between the LLRs transmitted from $\mathcal{O}_i^{-1}$ (resp. from $\mathcal{I}_{i+k}^{-1}$) to $\mathcal{O}_{i+k}^{-1}$ (resp. to $\mathcal{O}_i^{-1}$);
\item Same definitions hold for the $I_{e}^k(i^-)$ and $I_{a}^k(i^-)$ with respect to the $\mathcal{O}_{i}^{-1}$ and $\mathcal{O}_{i-k}^{-1}$.
\end{itemize}

Concerning the scheduling for spatially-coupled SC-TC decoding and analogously to BP for LDPC codes \cite{Urbanke_Richardson_MCT_Book}, we perform all inner updates (inner code pass) then all outer decoders updates (outer code pass) of each iteration the following mixtures rules:
\begin{itemize}
\item $I_{e}^k(i^+) = I_{e}(\mathcal{I}_i).b_k$ and $I_{a}(\mathcal{I}_i) = \sum I_{a}^k(i^+).b_k$ 
\item $I_{e}^k(i^-) = I_{e}(\mathcal{O}_i).b_k$ and $I_{a}(\mathcal{O}_i) = \sum I_{a}^k(i^-).b_k$
\item The \emph{a priori} MIs got from the added boundary nodes are equal to $1$.
\end{itemize}

The threshold of the spatially-coupled SC-TC is then defined as the lowest $E_b/N_0$ such as $I_{ap}(\mathcal{O}_i) \rightarrow 1, \forall i$.

\section{Numerical results}
\label{Section_Simulations}


In this section, we provide numerical results for various families of LDPC/GLDPC/DGLDPC and serially concatenated turbo codes. We choose BPSK, QPSK, $16$-QAM, and $64$-QAM modulation schemes for simulations. 
%
For simulations, we consider the following schemes:
\begin{itemize}
\item $\bm{\mathcal{S}_1}$: An outer rate $1/2$ systematic recursive $[5,7]$ convolutional code with an inner rate $1$ recursive accumulator
of transfer function $1/1+D$ with BPSK modulation 
\item $\bm{\mathcal{S}_2}$: Two serially concatenated rate-$1/2$ systematic recursive $[5,7]$ convolutional codes with $64$-QAM Gray mapping
\item $\bm{\mathcal{S}_3}$: An outer rate $1/2$ systematic recursive $[5,7]$ convolutional code with an inner rate $1$ recursive accumulator
of transfer function $1/1+D$ with $16$-QAM Gray mapping 
\item $\bm{\mathcal{S}_4}$: Two serially concatenated rate-$1/2$ systematic recursive $[5,7]$ convolutional codes with $16$-QAM SP mapping
%
%
\item $\bm{\mathcal{S}_5}$: $(4,8)$-regular LDPC code ensemble of rate $1/2$ with $64$-QAM Gray mapping
\item $\bm{\mathcal{S}_6}$: $(2,15)$-regular ensemble of design rate $7/15$ based on the Hamming($15,11$) component code designed in~\cite{Liva_Quasi_Cyclic_2008} with QPSK modulation
\item $\bm{\mathcal{S}_7}$: DGLDPC ensemble of rate $7/15$ from~\cite{Yige_DG_Globecomm_2006} with BPSK modulation. The structure of the 
generalized VNs and CNs for $\mathcal{S}_7$ is illustrated next. Suppose the generator matrices $G_1$ and $G_2$ are given by
\scriptsize
$G_1 = 
\begin{bmatrix}
1 & 0 & 0 & 1 & 1 & 0 \\
0 & 1 & 0 & 0 & 1 & 1  \\
0 & 0 & 1 & 1 & 0 & 1  
\end{bmatrix},
G_2 = 
\begin{bmatrix}
1 & 1 & 1 & 0 & 0 & 0 \\
0 & 1 & 1 & 1 & 0 & 0  \\
0 & 0 & 1 & 1 & 1 & 0 \\
0 & 0 & 0 & 1 & 1 & 1 
\end{bmatrix}.$
\normalsize
All VNs have degree $6$ and correspond to repetition codes of length $6$ ($69\%$ of all nodes), linear codes defined by $G_1$ ($1\%$), linear codes defined by $G_2$ ($22\%$) and single parity check codes of length $6$, denoted by SPC($6$), ($8\%$). All CNs nodes correspond to SPC($12$).
\item $\bm{\mathcal{S}_8}$: $(3,6)$-regular LDPC code ensemble of rate $1/2$ with $64$-QAM Natural mapping
\end{itemize}

\begin{figure*}[htbp]
\begin{center}
%
%
\input{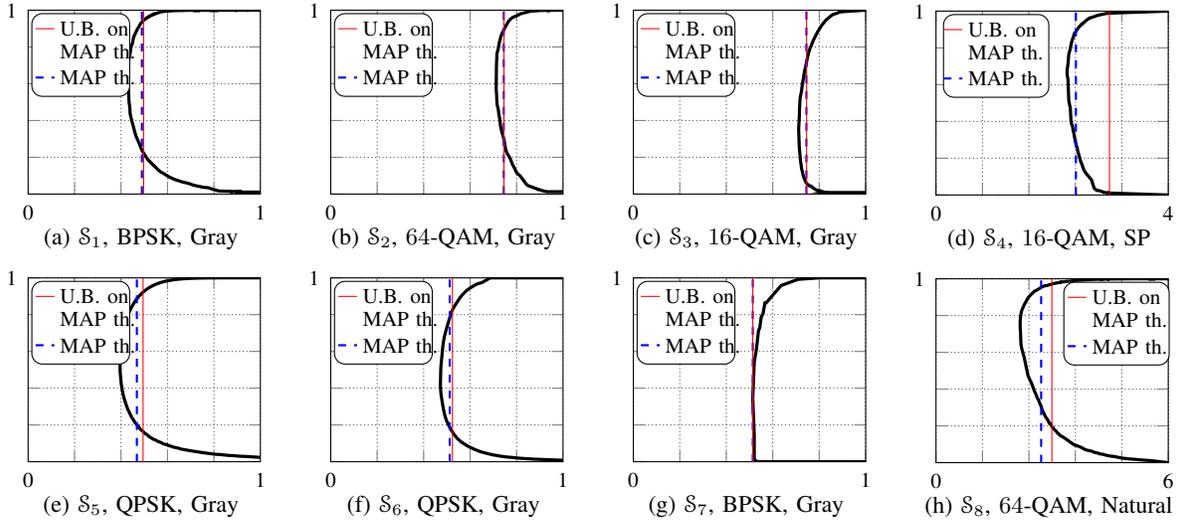}
%
\begin{tikzpicture}
\begin{axis}[axis_x_1_y_1_small]
\addplot [color = black, black, very thick] coordinates  {
( 1, 1.273095e-02)
(9.304355e-01, 1.273095e-02)
(9.145131e-01, 1.693733e-02)
(8.954719e-01, 2.507506e-02)
(8.912553e-01, 2.517245e-02)
(8.933813e-01, 2.523999e-02)
(8.890936e-01, 2.613720e-02)
(8.868959e-01, 2.838316e-02)
(8.846619e-01, 2.987927e-02)
(8.823912e-01, 3.058391e-02)
(8.800835e-01, 3.147876e-02)
(8.777386e-01, 3.240881e-02)
(8.753561e-01, 3.430075e-02)
(8.729359e-01, 3.488998e-02)
(8.704776e-01, 3.623163e-02)
(8.679811e-01, 3.736301e-02)
(8.654461e-01, 3.839754e-02)
(8.628725e-01, 4.040470e-02)
(8.602601e-01, 4.053503e-02)
(8.576088e-01, 4.175164e-02)
(8.549183e-01, 4.299672e-02)
(8.521887e-01, 4.417559e-02)
(8.494199e-01, 4.610953e-02)
(8.468316e-01, 4.783703e-02)
(8.439901e-01, 5.122626e-02)
(8.434173e-01, 5.190320e-02)
(8.425553e-01, 5.339929e-02)
(8.416899e-01, 5.395484e-02)
(8.167282e-01, 9.677907e-02)
(8.135117e-01, 1.007349e-01)
(8.102581e-01, 1.046579e-01)
(8.069673e-01, 1.079391e-01)
(8.036397e-01, 1.110153e-01)
(8.002753e-01, 1.170421e-01)
(7.968745e-01, 1.257004e-01)
(7.934374e-01, 1.342526e-01)
(7.899643e-01, 1.437817e-01)
(7.864555e-01, 1.518952e-01)
(7.829114e-01, 1.594718e-01)
(7.720702e-01, 1.949352e-01)
(7.646729e-01, 2.109027e-01)
(7.639259e-01, 2.124317e-01)
(7.631775e-01, 2.139545e-01)
(7.624279e-01, 2.154677e-01)
(7.616769e-01, 2.169682e-01)
(7.609246e-01, 2.184533e-01)
(7.601711e-01, 2.199209e-01)
(7.594162e-01, 2.225480e-01)
(7.586601e-01, 2.239697e-01)
(7.579027e-01, 2.253693e-01)
(7.571440e-01, 2.279021e-01)
(7.563840e-01, 2.303925e-01)
(7.556228e-01, 2.317091e-01)
(7.548603e-01, 2.352458e-01)
(7.540965e-01, 2.376218e-01)
(7.533315e-01, 2.421742e-01)
(7.525652e-01, 2.456010e-01)
(7.517976e-01, 2.511993e-01)
(7.510289e-01, 2.568066e-01)
(7.502589e-01, 2.635454e-01)
(7.494876e-01, 2.703628e-01)
(7.487151e-01, 2.772863e-01)
(7.479414e-01, 2.831975e-01)
(7.471665e-01, 2.880512e-01)
(7.463904e-01, 2.929646e-01)
(7.456131e-01, 2.967600e-01)
(7.448345e-01, 3.017664e-01)
(7.440548e-01, 3.044357e-01)
(7.432738e-01, 3.083101e-01)
(7.424917e-01, 3.122023e-01)
(7.417084e-01, 3.161046e-01)
(7.409239e-01, 3.187989e-01)
(7.401382e-01, 3.226887e-01)
(7.393514e-01, 3.265521e-01)
(7.385634e-01, 3.291760e-01)
(7.377742e-01, 3.329475e-01)
(7.369839e-01, 3.366540e-01)
(7.361924e-01, 3.402878e-01)
(7.353998e-01, 3.449967e-01)
(7.346061e-01, 3.484600e-01)
(7.338112e-01, 3.529712e-01)
(7.330152e-01, 3.573823e-01)
(7.322181e-01, 3.617032e-01)
(7.314199e-01, 3.670476e-01)
(7.306205e-01, 3.734236e-01)
(7.298201e-01, 3.808640e-01)
(7.290185e-01, 3.916329e-01)
(7.282159e-01, 4.003019e-01)
(7.274122e-01, 4.080104e-01)
(7.266074e-01, 4.147722e-01)
(7.258015e-01, 4.205737e-01)
(7.249946e-01, 4.253723e-01)
(7.241866e-01, 4.303150e-01)
(7.233775e-01, 4.366006e-01)
(7.225674e-01, 4.478996e-01)
(7.217563e-01, 4.555110e-01)
(7.209441e-01, 4.606120e-01)
(7.201309e-01, 4.656100e-01)
(7.193166e-01, 4.693000e-01)
(7.185014e-01, 4.739892e-01)
(7.176851e-01, 4.784778e-01)
(7.168678e-01, 4.838492e-01)
(7.160495e-01, 4.900574e-01)
(7.152302e-01, 5.022944e-01)
(7.144100e-01, 5.312309e-01)
(7.135887e-01, 5.468655e-01)
(7.123550e-01, 5.681060e-01)
(7.123550e-01, 5.969067e-01)
(7.135887e-01, 7.116331e-01)
(7.144100e-01, 7.302490e-01)
(7.152302e-01, 7.542032e-01)
(7.160495e-01, 7.662775e-01)
(7.168678e-01, 7.715819e-01)
(7.176851e-01, 7.756765e-01)
(7.185014e-01, 7.786387e-01)
(7.193166e-01, 7.815126e-01)
(7.201309e-01, 7.852794e-01)
(7.209441e-01, 7.899386e-01)
(7.217563e-01, 7.955496e-01)
(7.225674e-01, 8.022488e-01)
(7.233775e-01, 8.091202e-01)
(7.241866e-01, 8.140252e-01)
(7.249946e-01, 8.190089e-01)
(7.258015e-01, 8.240597e-01)
(7.266074e-01, 8.280508e-01)
(7.274122e-01, 8.320551e-01)
(7.282159e-01, 8.360573e-01)
(7.290185e-01, 8.400418e-01)
(7.298201e-01, 8.439930e-01)
(7.306205e-01, 8.478960e-01)
(7.314199e-01, 8.517375e-01)
(7.322181e-01, 8.555056e-01)
(7.330152e-01, 8.591910e-01)
(7.338112e-01, 8.617720e-01)
(7.346061e-01, 8.643061e-01)
(7.353998e-01, 8.667974e-01)
(7.361924e-01, 8.692506e-01)
(7.369839e-01, 8.716710e-01)
(7.377742e-01, 8.740639e-01)
(7.385634e-01, 8.754743e-01)
(7.393514e-01, 8.778361e-01)
(7.401382e-01, 8.801863e-01)
(7.409239e-01, 8.815842e-01)
(7.417084e-01, 8.839317e-01)
(7.424917e-01, 8.862828e-01)
(7.432738e-01, 8.877037e-01)
(7.440548e-01, 8.900781e-01)
(7.448345e-01, 8.915319e-01)
(7.456131e-01, 8.939438e-01)
(7.463904e-01, 8.963778e-01)
(7.471665e-01, 8.978976e-01)
(7.479414e-01, 9.003734e-01)
(7.487151e-01, 9.019316e-01)
(7.494876e-01, 9.035056e-01)
(7.502589e-01, 9.060296e-01)
(7.510289e-01, 9.076295e-01)
(7.517976e-01, 9.092390e-01)
(7.525652e-01, 9.108560e-01)
(7.533315e-01, 9.124785e-01)
(7.540965e-01, 9.141043e-01)
(7.548603e-01, 9.157313e-01)
(7.556228e-01, 9.182877e-01)
(7.563840e-01, 9.199093e-01)
(7.571440e-01, 9.215257e-01)
(7.579027e-01, 9.231349e-01)
(7.586601e-01, 9.247348e-01)
(7.594162e-01, 9.263236e-01)
(7.601711e-01, 9.278994e-01)
(7.609246e-01, 9.294602e-01)
(7.616769e-01, 9.310043e-01)
(7.624279e-01, 9.334356e-01)
(7.631775e-01, 9.349376e-01)
(7.639259e-01, 9.364177e-01)
(7.646729e-01, 9.378744e-01)
(7.720702e-01, 9.494578e-01)
(7.829114e-01, 9.603438e-01)
(7.864555e-01, 9.620969e-01)
(7.899643e-01, 9.645001e-01)
(7.934374e-01, 9.660584e-01)
(7.968745e-01, 9.676386e-01)
(8.002753e-01, 9.693189e-01)
(8.036397e-01, 9.711518e-01)
(8.069673e-01, 9.730848e-01)
(8.102581e-01, 9.758112e-01)
(8.135117e-01, 9.784688e-01)
(8.167282e-01, 9.801933e-01)
(8.416899e-01, 9.886051e-01)
(8.425553e-01, 9.888043e-01)
(8.434173e-01, 9.890024e-01)
(8.439901e-01, 9.898100e-01)
(8.468316e-01, 9.904212e-01)
(8.494199e-01, 9.916136e-01)
(8.521887e-01, 9.920921e-01)
(8.549183e-01, 9.931227e-01)
(8.576088e-01, 9.934797e-01)
(8.602601e-01, 9.943937e-01)
(8.628725e-01, 9.946589e-01)
(8.654461e-01, 9.948849e-01)
(8.679811e-01, 9.950767e-01)
(8.704776e-01, 9.952392e-01)
(8.729359e-01, 9.953775e-01)
(8.753561e-01, 9.961808e-01)
(8.777386e-01, 9.962789e-01)
(8.800835e-01, 9.963633e-01)
(8.823912e-01, 9.964420e-01)
(8.846619e-01, 9.965228e-01)
(8.868959e-01, 9.966132e-01)
(8.890936e-01, 9.967179e-01)
(8.912553e-01, 9.975989e-01)
(8.933813e-01, 9.977592e-01)
(8.954719e-01, 9.979266e-01)
(9.145131e-01, 9.986691e-01)
(9.436391e-01, 9.990614e-01)
(9.304355e-01, 9.991052e-01)
( 1,  1)
};
%
\addplot [color = blue, blue, thick, dashed] coordinates {(0.7448, 0) (0.7448, 1)};
\addplot [color = red, red] coordinates {(0.7456, 0) (0.7456, 1)};
%
\end{axis}
\draw [rounded corners,fill=white] (0.05,1.7+0.7) rectangle (1.35,0.6+0.7);
\draw [red] (0.1,1.5+0.7) -- (0.35,1.5+0.7);
\node [right] at (0.28,1.5+0.7) {\scriptsize{U.B.~on}};
\node [right] at (0.28,1.5-0.3+0.7) {\scriptsize{MAP th.}};
\draw [blue, thick, dashed] (0.1,1.5-0.65+0.7) -- (0.35,1.5-0.65+0.7);
\node [right] at (0.28,1.5-0.65+0.7) {\scriptsize{MAP th.}};
\node [below] at (1.5,-0.3) {\footnotesize{(b) $\mathcal{S}_2$, $64$-QAM, Gray}};
\end{tikzpicture}
%
\begin{tikzpicture}
\begin{axis}[axis_x_1_y_1_small]
\addplot [color = black, black, very thick] coordinates  {
( 1, 8.338051e-03)
(8.264599e-01, 8.338051e-03)
(8.296242e-01, 8.450958e-03)
(8.199993e-01, 9.379525e-03)
(8.232517e-01, 1.007681e-02)
(8.167026e-01, 1.234398e-02)
(8.133615e-01, 1.310657e-02)
(8.099757e-01, 1.564931e-02)
(8.030698e-01, 1.804947e-02)
(7.923738e-01, 2.089502e-02)
(7.850177e-01, 2.414617e-02)
(7.736451e-01, 3.289818e-02)
(7.717103e-01, 3.480400e-02)
(7.697642e-01, 3.924313e-02)
(7.678069e-01, 4.101245e-02)
(7.658384e-01, 4.146499e-02)
(7.638587e-01, 4.314831e-02)
(7.618677e-01, 4.355554e-02)
(7.598657e-01, 4.395585e-02)
(7.578524e-01, 4.554024e-02)
(7.558280e-01, 4.592213e-02)
(7.537925e-01, 4.863275e-02)
(7.517459e-01, 5.132279e-02)
(7.496882e-01, 5.283476e-02)
(7.476195e-01, 5.548017e-02)
(7.455398e-01, 5.808206e-02)
(7.434490e-01, 6.176864e-02)
(7.413473e-01, 6.765778e-02)
(7.392346e-01, 7.456869e-02)
(7.371111e-01, 8.017591e-02)
(7.349766e-01, 8.676309e-02)
(7.328313e-01, 9.320090e-02)
(7.306752e-01, 1.029089e-01)
(7.285083e-01, 1.160083e-01)
(7.263307e-01, 1.280860e-01)
(7.241423e-01, 1.378497e-01)
(7.219433e-01, 1.604583e-01)
(7.197337e-01, 1.770875e-01)
(7.175135e-01, 2.103037e-01)
(7.152827e-01, 2.384161e-01)
(7.143875e-01, 2.520841e-01)
(7.116917e-01, 3.448811e-01)
(7.116917e-01, 3.584773e-01)
(7.143875e-01, 4.201682e-01)
(7.152827e-01, 4.613252e-01)
(7.175135e-01, 5.165463e-01)
(7.197337e-01, 5.357329e-01)
(7.219433e-01, 5.475967e-01)
(7.241423e-01, 5.621293e-01)
(7.263307e-01, 5.857211e-01)
(7.285083e-01, 6.055110e-01)
(7.306752e-01, 6.201265e-01)
(7.328313e-01, 6.354653e-01)
(7.349766e-01, 6.532410e-01)
(7.371111e-01, 6.683000e-01)
(7.392346e-01, 6.808959e-01)
(7.413473e-01, 6.922432e-01)
(7.434490e-01, 7.037241e-01)
(7.455398e-01, 7.167472e-01)
(7.476195e-01, 7.307968e-01)
(7.496882e-01, 7.440241e-01)
(7.517459e-01, 7.552016e-01)
(7.537925e-01, 7.655390e-01)
(7.558280e-01, 7.743976e-01)
(7.578524e-01, 7.832298e-01)
(7.598657e-01, 7.903289e-01)
(7.618677e-01, 7.964548e-01)
(7.638587e-01, 8.023713e-01)
(7.658384e-01, 8.081106e-01)
(7.678069e-01, 8.137517e-01)
(7.697642e-01, 8.201260e-01)
(7.717103e-01, 8.265843e-01)
(7.736451e-01, 8.338627e-01)
(7.850177e-01, 8.621840e-01)
(7.923738e-01, 8.802458e-01)
(8.030698e-01, 9.025817e-01)
(8.099757e-01, 9.161454e-01)
(8.133615e-01, 9.216850e-01)
(8.167026e-01, 9.270504e-01)
(8.199993e-01, 9.323287e-01)
(8.232517e-01, 9.375424e-01)
(8.264599e-01, 9.426515e-01)
(8.296242e-01, 9.468992e-01)
(8.327447e-01, 9.502971e-01)
(8.358217e-01, 9.535915e-01)
(8.388554e-01, 9.568657e-01)
(8.418460e-01, 9.602200e-01)
(8.447938e-01, 9.630568e-01)
(8.476991e-01, 9.660068e-01)
(8.505622e-01, 9.689785e-01)
(8.533832e-01, 9.718677e-01)
(8.560010e-01, 9.739047e-01)
(8.587402e-01, 9.763449e-01)
(8.614368e-01, 9.772635e-01)
(8.640911e-01, 9.787010e-01)
(8.667034e-01, 9.794337e-01)
(8.692739e-01, 9.808363e-01)
(8.718030e-01, 9.816656e-01)
(8.742911e-01, 9.825680e-01)
(8.767385e-01, 9.835254e-01)
(8.791456e-01, 9.851927e-01)
(8.815127e-01, 9.862073e-01)
(8.838403e-01, 9.879065e-01)
(8.861288e-01, 9.889213e-01)
(8.883785e-01, 9.892372e-01)
(8.905899e-01, 9.902273e-01)
(8.927634e-01, 9.911966e-01)
( 1,  1)
};
%
\addplot [color = blue, blue, thick, dashed] coordinates {(0.7448, 0) (0.7448, 1)};
\addplot [color = red, red] coordinates {(0.7456, 0) (0.7456, 1)};
%
\end{axis}
\draw [rounded corners,fill=white] (0.05,1.7+0.7) rectangle (1.35,0.6+0.7);
\draw [red] (0.1,1.5+0.7) -- (0.35,1.5+0.7);
\node [right] at (0.28,1.5+0.7) {\scriptsize{U.B.~on}};
\node [right] at (0.28,1.5-0.3+0.7) {\scriptsize{MAP th.}};
\draw [blue, thick, dashed] (0.1,1.5-0.65+0.7) -- (0.35,1.5-0.65+0.7);
\node [right] at (0.28,1.5-0.65+0.7) {\scriptsize{MAP th.}};
\node [below] at (1.5,-0.3) {\footnotesize{(c) $\mathcal{S}_3$, $16$-QAM, Gray}};
\end{tikzpicture}
%
\begin{tikzpicture}
\begin{axis}[axis_x_4_y_1_small]
\addplot [color = black, black, very thick] coordinates  {
( 4,  0)
(2.954554e+00, 1.089143e-02)
(2.987413e+00, 1.103087e-02)
(2.920248e+00, 1.255732e-02)
(2.884419e+00, 1.488929e-02)
(2.846984e+00, 2.037159e-02)
(2.807857e+00, 2.374462e-02)
(2.766946e+00, 2.515676e-02)
(2.724150e+00, 3.665732e-02)
(2.679364e+00, 8.042499e-02)
(2.660868e+00, 8.911946e-02)
(2.642027e+00, 9.458757e-02)
(2.622834e+00, 1.037145e-01)
(2.603281e+00, 1.093122e-01)
(2.583360e+00, 1.184795e-01)
(2.573258e+00, 1.218331e-01)
(2.563061e+00, 1.269242e-01)
(2.552768e+00, 1.337283e-01)
(2.542377e+00, 1.385117e-01)
(2.531887e+00, 1.489478e-01)
(2.521297e+00, 1.582760e-01)
(2.510606e+00, 1.694864e-01)
(2.499813e+00, 1.749178e-01)
(2.488916e+00, 1.853981e-01)
(2.477914e+00, 1.986518e-01)
(2.466807e+00, 2.107328e-01)
(2.455591e+00, 2.240008e-01)
(2.444267e+00, 2.331782e-01)
(2.432833e+00, 2.547099e-01)
(2.421287e+00, 2.643485e-01)
(2.409629e+00, 2.786659e-01)
(2.397856e+00, 2.910775e-01)
(2.385967e+00, 3.148364e-01)
(2.373962e+00, 3.345177e-01)
(2.361837e+00, 3.495777e-01)
(2.349593e+00, 3.773403e-01)
(2.337227e+00, 3.821808e-01)
(2.324737e+00, 4.048985e-01)
(2.312122e+00, 4.751591e-01)
(2.299381e+00, 4.926640e-01)
(2.286512e+00, 5.128774e-01)
(2.273513e+00, 6.267329e-01)
(2.260382e+00, 6.507737e-01)
(2.259062e+00, 6.562533e-01)
(2.257740e+00, 6.633284e-01)
(2.257079e+00, 6.707564e-01)
(2.257079e+00, 6.720955e-01)
(2.257740e+00, 6.745059e-01)
(2.259062e+00, 6.800521e-01)
(2.260382e+00, 6.846524e-01)
(2.273513e+00, 7.139795e-01)
(2.286512e+00, 7.648048e-01)
(2.299381e+00, 7.802669e-01)
(2.312122e+00, 7.958310e-01)
(2.324737e+00, 8.157201e-01)
(2.337227e+00, 8.337006e-01)
(2.349593e+00, 8.462610e-01)
(2.361837e+00, 8.562820e-01)
(2.373962e+00, 8.683912e-01)
(2.385967e+00, 8.809551e-01)
(2.397856e+00, 8.836833e-01)
(2.409629e+00, 8.889306e-01)
(2.421287e+00, 8.953398e-01)
(2.432833e+00, 9.001801e-01)
(2.444267e+00, 9.055243e-01)
(2.455591e+00, 9.121255e-01)
(2.466807e+00, 9.151645e-01)
(2.477914e+00, 9.185103e-01)
(2.488916e+00, 9.219536e-01)
(2.499813e+00, 9.272010e-01)
(2.510606e+00, 9.275036e-01)
(2.521297e+00, 9.305626e-01)
(2.531887e+00, 9.351659e-01)
(2.542377e+00, 9.366523e-01)
(2.552768e+00, 9.399110e-01)
(2.563061e+00, 9.431369e-01)
(2.573258e+00, 9.457069e-01)
(2.583360e+00, 9.485154e-01)
(2.603281e+00, 9.534124e-01)
(2.622834e+00, 9.572543e-01)
(2.642027e+00, 9.621916e-01)
(2.660868e+00, 9.626708e-01)
(2.679364e+00, 9.661845e-01)
(2.724150e+00, 9.698855e-01)
(2.766946e+00, 9.753893e-01)
(2.807857e+00, 9.782782e-01)
(2.846984e+00, 9.817211e-01)
(2.884419e+00, 9.838777e-01)
(2.920248e+00, 9.862177e-01)
(2.954554e+00, 9.881767e-01)
(2.987413e+00, 9.888276e-01)
( 4,  1)
};
\addplot [color = blue, blue, thick, dashed] coordinates {(2.4096, 0) (2.4096, 1)};
\addplot [color = red, red] coordinates {(2.9874, 0) (2.9874, 1)};
%
\end{axis} 
\draw [rounded corners,fill=white] (0.05+1.65-1.6,1.7+0.7) rectangle (1.35+1.75-1.6,0.6+0.7);
\draw [red] (0.1,1.5+0.7) -- (0.35,1.5+0.7);
\node [right] at (0.28,1.5+0.7) {\scriptsize{U.B.~on}};
\node [right] at (0.28,1.5-0.3+0.7) {\scriptsize{MAP th.}};
\draw [blue, thick, dashed] (0.1,1.5-0.65+0.7) -- (0.35,1.5-0.65+0.7);
\node [right] at (0.28,1.5-0.65+0.7) {\scriptsize{MAP th.}};
\node [below] at (1.5,-0.3) {\footnotesize{(d) $\mathcal{S}_4$, $16$-QAM, SP}};
\end{tikzpicture}
%
%
\begin{tikzpicture}
\begin{axis}[axis_x_1_y_1_small]
\addplot [color=black,black, very thick] coordinates  {
(9.995785e-01, 2.406560e-02)
(9.932469e-01, 2.415042e-02)
(9.647782e-01, 2.654633e-02)
(9.309995e-01, 2.920539e-02)
(9.179962e-01, 2.940965e-02)
(9.052540e-01, 3.075248e-02)
(8.927577e-01, 3.213346e-02)
(8.804941e-01, 3.355214e-02)
(8.684521e-01, 3.376218e-02)
(8.566219e-01, 3.522105e-02)
(8.449947e-01, 3.671642e-02)
(8.335631e-01, 3.824788e-02)
(8.223200e-01, 3.981507e-02)
(8.112592e-01, 4.003135e-02)
(8.003753e-01, 4.163645e-02)
(7.896629e-01, 4.327663e-02)
(7.791174e-01, 4.495164e-02)
(7.687345e-01, 4.666126e-02)
(7.585100e-01, 4.840530e-02)
(7.484403e-01, 5.018358e-02)
(7.385218e-01, 5.199592e-02)
(7.287513e-01, 5.384212e-02)
(7.191257e-01, 5.572202e-02)
(7.096421e-01, 5.763542e-02)
(7.002979e-01, 5.958211e-02)
(6.910904e-01, 6.156188e-02)
(6.820171e-01, 6.357450e-02)
(6.730759e-01, 6.561973e-02)
(6.642645e-01, 6.769731e-02)
(6.555808e-01, 6.980697e-02)
(6.470228e-01, 7.386130e-02)
(6.385886e-01, 7.606312e-02)
(6.309364e-01, 7.829580e-02)
(6.227677e-01, 8.055900e-02)
(6.147252e-01, 8.490097e-02)
(6.068060e-01, 8.725150e-02)
(5.990073e-01, 9.175515e-02)
(5.913265e-01, 9.418976e-02)
(5.837611e-01, 9.665224e-02)
(5.763084e-01, 1.013640e-01)
(5.689662e-01, 1.061717e-01)
(5.617323e-01, 1.087637e-01)
(5.546043e-01, 1.137129e-01)
(5.475803e-01, 1.187501e-01)
(5.406581e-01, 1.214611e-01)
(5.338358e-01, 1.266274e-01)
(5.271114e-01, 1.318733e-01)
(5.204832e-01, 1.371955e-01)
(5.139493e-01, 1.451408e-01)
(5.075079e-01, 1.506359e-01)
(5.011575e-01, 1.561952e-01)
(4.948963e-01, 1.644637e-01)
(4.887227e-01, 1.701657e-01)
(4.826353e-01, 1.786257e-01)
(4.766325e-01, 1.886094e-01)
(4.707128e-01, 1.975212e-01)
(4.648747e-01, 2.065713e-01)
(4.591170e-01, 2.157512e-01)
(4.534383e-01, 2.280384e-01)
(4.478371e-01, 2.374862e-01)
(4.423123e-01, 2.500954e-01)
(4.368626e-01, 2.659633e-01)
(4.314867e-01, 2.788954e-01)
(4.261834e-01, 2.951041e-01)
(4.209516e-01, 3.146418e-01)
(4.157901e-01, 3.343194e-01)
(4.106978e-01, 3.573000e-01)
(4.056736e-01, 3.867430e-01)
(4.007163e-01, 4.256405e-01)
(3.958251e-01, 4.949694e-01)
(3.958251e-01, 5.832341e-01)
(4.007163e-01, 6.547953e-01)
(4.056736e-01, 6.932751e-01)
(4.106978e-01, 7.245166e-01)
(4.157901e-01, 7.514994e-01)
(4.209516e-01, 7.727795e-01)
(4.261834e-01, 7.909248e-01)
(4.314867e-01, 8.038386e-01)
(4.368626e-01, 8.181717e-01)
(4.423123e-01, 8.332932e-01)
(4.478371e-01, 8.461185e-01)
(4.534383e-01, 8.582415e-01)
(4.591170e-01, 8.696735e-01)
(4.648747e-01, 8.804285e-01)
(4.707128e-01, 8.894053e-01)
(4.766325e-01, 8.989288e-01)
(4.826353e-01, 9.068441e-01)
(4.887227e-01, 9.152067e-01)
(4.948963e-01, 9.221275e-01)
(5.011575e-01, 9.286064e-01)
(5.075079e-01, 9.354097e-01)
(5.139493e-01, 9.410049e-01)
(5.204832e-01, 9.462122e-01)
(5.271114e-01, 9.510494e-01)
(5.338358e-01, 9.555343e-01)
(5.406581e-01, 9.601960e-01)
(5.475803e-01, 9.639898e-01)
(5.546043e-01, 9.674861e-01)
(5.617323e-01, 9.707019e-01)
(5.689662e-01, 9.740162e-01)
(5.763084e-01, 9.766897e-01)
(5.837611e-01, 9.791332e-01)
(5.913265e-01, 9.816346e-01)
(5.990073e-01, 9.836383e-01)
(6.068060e-01, 9.856796e-01)
(6.147252e-01, 9.873066e-01)
(6.227677e-01, 9.889559e-01)
(6.309364e-01, 9.902635e-01)
(6.385886e-01, 9.915820e-01)
(6.470228e-01, 9.926217e-01)
(6.555808e-01, 9.936642e-01)
(6.642645e-01, 9.945803e-01)
(6.730759e-01, 9.954675e-01)
(6.820171e-01, 9.961564e-01)
(6.910904e-01, 9.967550e-01)
(7.002979e-01, 9.973278e-01)
(7.096421e-01, 9.977667e-01)
(7.191257e-01, 9.981828e-01)
(7.287513e-01, 9.985315e-01)
(7.385218e-01, 9.987946e-01)
(7.484403e-01, 9.990399e-01)
(7.585100e-01, 9.992602e-01)
(7.687345e-01, 9.994216e-01)
(7.791174e-01, 9.995523e-01)
(7.896629e-01, 9.996667e-01)
(8.003753e-01, 9.997554e-01)
(8.112592e-01, 9.998234e-01)
(8.223200e-01, 9.998748e-01)
(8.335631e-01, 9.999131e-01)
(8.449947e-01, 9.999410e-01)
(8.566219e-01, 9.999625e-01)
(8.684521e-01, 9.999771e-01)
(8.804941e-01, 9.999866e-01)
(8.927577e-01, 1.000000e+00)
(9.052540e-01, 1.000000e+00)
(9.179962e-01, 1.000000e+00)
(9.309995e-01, 1.000000e+00)
(9.647782e-01, 1.000000e+00)
(9.932469e-01, 1.000000e+00)
(9.995785e-01, 1.000000e+00)
( 1,  1)
};
%
\addplot [color = blue, blue, thick, dashed] coordinates {(0.4682, 0) (0.4682, 1)};
\addplot [color = red, red] coordinates {(0.4949, 0) (0.4949, 1)};
%
\end{axis}
\draw [rounded corners,fill=white] (0.05,1.7+0.7) rectangle (1.35,0.6+0.7);
\draw [red] (0.1,1.5+0.7) -- (0.35,1.5+0.7);
\node [right] at (0.28,1.5+0.7) {\scriptsize{U.B.~on}};
\node [right] at (0.28,1.5-0.3+0.7) {\scriptsize{MAP th.}};
\draw [blue, thick, dashed] (0.1,1.5-0.65+0.7) -- (0.35,1.5-0.65+0.7);
\node [right] at (0.28,1.5-0.65+0.7) {\scriptsize{MAP th.}};
\node [below] at (1.5,-0.3) {\footnotesize{(e) $\mathcal{S}_5$, QPSK, Gray}};
\end{tikzpicture}
\begin{tikzpicture}
\begin{axis}[axis_x_1_y_1_small]
\addplot [color = black, black, very thick] coordinates  {
%
( 1, 0)
(9.932469e-01, 8.336629e-03)
(9.788574e-01, 9.078124e-03)
(9.309995e-01, 1.122277e-02)
(9.147868e-01, 1.214991e-02)
(8.989760e-01, 1.312496e-02)
(8.835388e-01, 1.368513e-02)
(8.684521e-01, 1.473550e-02)
(8.536964e-01, 1.583754e-02)
(8.392549e-01, 1.699288e-02)
(8.251134e-01, 1.820312e-02)
(8.112592e-01, 1.946972e-02)
(8.003753e-01, 2.076718e-02)
(7.949979e-01, 2.143694e-02)
(7.896629e-01, 2.212083e-02)
(7.843696e-01, 2.281891e-02)
(7.791174e-01, 2.353121e-02)
(7.739059e-01, 2.425777e-02)
(7.687345e-01, 2.431718e-02)
(7.636027e-01, 2.505923e-02)
(7.585100e-01, 2.581553e-02)
(7.534560e-01, 2.658615e-02)
(7.484403e-01, 2.737116e-02)
(7.434624e-01, 2.817061e-02)
(7.385218e-01, 2.898457e-02)
(7.336183e-01, 2.981306e-02)
(7.287513e-01, 3.065613e-02)
(7.239206e-01, 3.151379e-02)
(7.191257e-01, 3.320018e-02)
(7.143664e-01, 3.410052e-02)
(7.096421e-01, 3.501548e-02)
(7.049528e-01, 3.594506e-02)
(7.002979e-01, 3.688924e-02)
(6.956772e-01, 3.784799e-02)
(6.910904e-01, 3.972763e-02)
(6.865371e-01, 4.072878e-02)
(6.512862e-01, 5.239722e-02)
(6.470228e-01, 5.356869e-02)
(6.427903e-01, 5.585906e-02)
(6.385886e-01, 5.707863e-02)
(6.350690e-01, 5.946534e-02)
(6.309364e-01, 6.192022e-02)
(6.268361e-01, 6.322929e-02)
(6.227677e-01, 6.579589e-02)
(6.187308e-01, 6.716543e-02)
(6.147252e-01, 6.985419e-02)
(6.107503e-01, 7.262963e-02)
(6.068060e-01, 7.411198e-02)
(6.028918e-01, 7.702904e-02)
(5.990073e-01, 7.858755e-02)
(5.951524e-01, 8.165906e-02)
(5.913265e-01, 8.483514e-02)
(5.875295e-01, 8.652776e-02)
(5.837611e-01, 8.985513e-02)
(5.800208e-01, 9.327651e-02)
(5.763084e-01, 9.509128e-02)
(5.726237e-01, 9.864541e-02)
(5.689662e-01, 1.022795e-01)
(5.653359e-01, 1.041988e-01)
(5.617323e-01, 1.079420e-01)
(5.581552e-01, 1.117475e-01)
(5.546043e-01, 1.156074e-01)
(5.510794e-01, 1.195132e-01)
(5.475803e-01, 1.234558e-01)
(5.441066e-01, 1.274259e-01)
(5.406581e-01, 1.333425e-01)
(5.372346e-01, 1.373816e-01)
(5.338358e-01, 1.434218e-01)
(5.304615e-01, 1.495613e-01)
(5.271114e-01, 1.558224e-01)
(5.237854e-01, 1.622291e-01)
(5.204832e-01, 1.688074e-01)
(5.172045e-01, 1.755855e-01)
(5.139493e-01, 1.825903e-01)
(5.123303e-01, 1.873505e-01)
(5.107171e-01, 1.898179e-01)
(5.091097e-01, 1.947113e-01)
(5.075079e-01, 1.996824e-01)
(5.059119e-01, 2.022491e-01)
(5.043215e-01, 2.073204e-01)
(5.027367e-01, 2.124460e-01)
(5.011575e-01, 2.150826e-01)
(4.995839e-01, 2.202711e-01)
(4.980158e-01, 2.254876e-01)
(4.964533e-01, 2.307209e-01)
(4.948963e-01, 2.359593e-01)
(4.933447e-01, 2.437402e-01)
(4.917987e-01, 2.489480e-01)
(4.902580e-01, 2.566847e-01)
(4.887227e-01, 2.644274e-01)
(4.871929e-01, 2.747675e-01)
(4.856684e-01, 2.826379e-01)
(4.841492e-01, 2.932614e-01)
(4.826353e-01, 3.041567e-01)
(4.811267e-01, 3.153980e-01)
(4.796234e-01, 3.269624e-01)
(4.781253e-01, 3.387718e-01)
(4.766325e-01, 3.507434e-01)
(4.751448e-01, 3.627895e-01)
(4.736623e-01, 3.807252e-01)
(4.721850e-01, 4.155041e-01)
(4.721850e-01, 4.211492e-01)
(4.736623e-01, 5.051555e-01)
(4.751448e-01, 5.299095e-01)
(4.766325e-01, 5.514596e-01)
(4.781253e-01, 5.702239e-01)
(4.796234e-01, 5.891887e-01)
(4.811267e-01, 6.085073e-01)
(4.826353e-01, 6.278490e-01)
(4.841492e-01, 6.413741e-01)
(4.856684e-01, 6.544886e-01)
(4.871929e-01, 6.670291e-01)
(4.887227e-01, 6.766216e-01)
(4.902580e-01, 6.858987e-01)
(4.917987e-01, 6.926435e-01)
(4.933447e-01, 7.014861e-01)
(4.948963e-01, 7.079634e-01)
(4.964533e-01, 7.143627e-01)
(4.980158e-01, 7.228530e-01)
(4.995839e-01, 7.291372e-01)
(5.011575e-01, 7.353911e-01)
(5.027367e-01, 7.416258e-01)
(5.043215e-01, 7.499716e-01)
(5.059119e-01, 7.562056e-01)
(5.075079e-01, 7.624592e-01)
(5.091097e-01, 7.708874e-01)
(5.107171e-01, 7.771988e-01)
(5.123303e-01, 7.834906e-01)
(5.139493e-01, 7.918254e-01)
(5.172045e-01, 8.038876e-01)
(5.204832e-01, 8.152216e-01)
(5.237854e-01, 8.255663e-01)
(5.271114e-01, 8.334546e-01)
(5.304615e-01, 8.408579e-01)
(5.338358e-01, 8.479186e-01)
(5.372346e-01, 8.547825e-01)
(5.406581e-01, 8.615686e-01)
(5.441066e-01, 8.682760e-01)
(5.475803e-01, 8.748729e-01)
(5.510794e-01, 8.826309e-01)
(5.546043e-01, 8.888682e-01)
(5.581552e-01, 8.948975e-01)
(5.617323e-01, 9.007124e-01)
(5.653359e-01, 9.063093e-01)
(5.689662e-01, 9.105995e-01)
(5.726237e-01, 9.157958e-01)
(5.763084e-01, 9.207621e-01)
(5.800208e-01, 9.254870e-01)
(5.837611e-01, 9.299572e-01)
(5.875295e-01, 9.341593e-01)
(5.913265e-01, 9.372971e-01)
(5.951524e-01, 9.409906e-01)
(5.990073e-01, 9.444459e-01)
(6.028918e-01, 9.470675e-01)
(6.068060e-01, 9.502766e-01)
(6.107503e-01, 9.534517e-01)
(6.147252e-01, 9.559988e-01)
(6.187308e-01, 9.592091e-01)
(6.227677e-01, 9.622998e-01)
(6.268361e-01, 9.651320e-01)
(6.309364e-01, 9.671036e-01)
(6.350690e-01, 9.691156e-01)
(6.385886e-01, 9.703331e-01)
(6.427903e-01, 9.713951e-01)
(6.470228e-01, 9.728218e-01)
(6.512862e-01, 9.747080e-01)
(6.865371e-01, 1.000000e+00)
(6.910904e-01, 1.000000e+00)
(6.956772e-01, 1.000000e+00)
(7.002979e-01, 1.000000e+00)
(7.049528e-01, 1.000000e+00)
(7.096421e-01, 1.000000e+00)
(7.143664e-01, 1.000000e+00)
(7.191257e-01, 1.000000e+00)
(7.239206e-01, 1.000000e+00)
(7.287513e-01, 1.000000e+00)
(7.336183e-01, 1.000000e+00)
(7.385218e-01, 1.000000e+00)
(7.434624e-01, 1.000000e+00)
(7.484403e-01, 1.000000e+00)
(7.534560e-01, 1.000000e+00)
(7.585100e-01, 1.000000e+00)
(7.636027e-01, 1.000000e+00)
(7.687345e-01, 1.000000e+00)
(7.739059e-01, 1.000000e+00)
(7.791174e-01, 1.000000e+00)
(7.843696e-01, 1.000000e+00)
(7.896629e-01, 1.000000e+00)
(7.949979e-01, 1.000000e+00)
(8.003753e-01, 1.000000e+00)
(8.112592e-01, 1.000000e+00)
(8.251134e-01, 1.000000e+00)
(8.392549e-01, 1.000000e+00)
(8.536964e-01, 1.000000e+00)
(8.684521e-01, 1.000000e+00)
(8.835388e-01, 1.000000e+00)
(8.989760e-01, 1.000000e+00)
(9.147868e-01, 1.000000e+00)
(9.309995e-01, 1.000000e+00)
(9.788574e-01, 1.000000e+00)
(9.932469e-01, 1.000000e+00)
( 1,  1)
};
%
\addplot [color = blue, blue, thick, dashed] coordinates {(0.5123, 0) (0.5123, 1)};
\addplot [color = red, red] coordinates {(0.5238, 0) (0.5238, 1)};
\end{axis}
\draw [rounded corners,fill=white] (0.05,1.7+0.7) rectangle (1.35,0.6+0.7);
\draw [red] (0.1,1.5+0.7) -- (0.35,1.5+0.7);
\node [right] at (0.28,1.5+0.7) {\scriptsize{U.B.~on}};
\node [right] at (0.28,1.5-0.3+0.7) {\scriptsize{MAP th.}};
\draw [blue, thick, dashed] (0.1,1.5-0.65+0.7) -- (0.35,1.5-0.65+0.7);
\node [right] at (0.28,1.5-0.65+0.7) {\scriptsize{MAP th.}};
\node [below] at (1.5,-0.3) {\footnotesize{(f) $\mathcal{S}_6$, QPSK, Gray}};
\end{tikzpicture}
\begin{tikzpicture}
\begin{axis}[axis_x_1_y_1_small]
\addplot [color = black, black, very thick] coordinates  {
(1,0)
(0.5445, 1.2912e-04)
(0.5387, 1.5841e-04)
(0.5298, 0.0014)
(0.5204, 0.0152)
(0.5204, 0.1497)
(0.5140,0.3439)
(0.5140,  0.3439)
(0.5204,  0.5420)
(0.5298,  0.6113)
(0.5387,  0.7123)
(0.5445,  0.7305)
(0.5502,  0.7862)
(0.5584,  0.8056)
(0.5663,  0.8638)
(0.5714,  0.8638)
(0.6362,  0.9585)
(0.7095,  0.9921)
(0.8,  1)
(1,  1)
};
%
\addplot [color = blue, blue, thick, dashed] coordinates {(0.514, 0) (0.514, 1)};
\addplot [color = red, red] coordinates {(0.514, 0) (0.514, 1)};
\end{axis}
\draw [rounded corners,fill=white] (0.05,1.7+0.7) rectangle (1.35,0.6+0.7);
\draw [red] (0.1,1.5+0.7) -- (0.35,1.5+0.7);
\node [right] at (0.28,1.5+0.7) {\scriptsize{U.B.~on}};
\node [right] at (0.28,1.5-0.3+0.7) {\scriptsize{MAP th.}};
\draw [blue, thick, dashed] (0.1,1.5-0.65+0.7) -- (0.35,1.5-0.65+0.7);
\node [right] at (0.28,1.5-0.65+0.7) {\scriptsize{MAP th.}};
\node [below] at (1.5,-0.3) {\footnotesize{(g) $\mathcal{S}_7$, BPSK, Gray}};
\end{tikzpicture}
\begin{tikzpicture}
\begin{axis}[axis_x_6_y_1_small]
\addplot [color = black, black, very thick] coordinates  {
( 6,  0)
(5.667603e+00, 8.244381e-03)
(5.590990e+00, 1.035355e-02)
(5.486291e+00, 1.192758e-02)
(5.339129e+00, 1.433226e-02)
(5.242727e+00, 1.585471e-02)
(5.125722e+00, 1.863879e-02)
(4.982471e+00, 2.050915e-02)
(4.805462e+00, 2.487739e-02)
(4.584479e+00, 3.428493e-02)
(4.533837e+00, 3.497482e-02)
(4.480739e+00, 3.677978e-02)
(4.425032e+00, 3.911482e-02)
(4.366547e+00, 4.200836e-02)
(4.305101e+00, 4.581840e-02)
(4.240494e+00, 4.733486e-02)
(4.172502e+00, 5.276903e-02)
(4.100883e+00, 5.539384e-02)
(4.025361e+00, 6.216819e-02)
(3.945636e+00, 6.593510e-02)
(3.861365e+00, 7.013017e-02)
(3.772165e+00, 8.080704e-02)
(3.677604e+00, 8.951753e-02)
(3.577188e+00, 9.607929e-02)
(3.470352e+00, 1.111197e-01)
(3.356447e+00, 1.291078e-01)
(3.234719e+00, 1.450069e-01)
(3.209360e+00, 1.467580e-01)
(3.183645e+00, 1.525518e-01)
(3.157567e+00, 1.591473e-01)
(3.131117e+00, 1.622905e-01)
(3.104288e+00, 1.720030e-01)
(3.077069e+00, 1.760151e-01)
(3.049452e+00, 1.810771e-01)
(3.021428e+00, 1.902991e-01)
(2.992986e+00, 1.952328e-01)
(2.964117e+00, 2.111562e-01)
(2.934810e+00, 2.122473e-01)
(2.905053e+00, 2.240901e-01)
(2.874837e+00, 2.366575e-01)
(2.844148e+00, 2.444778e-01)
(2.812975e+00, 2.561901e-01)
(2.781304e+00, 2.707952e-01)
(2.749124e+00, 2.833484e-01)
(2.716420e+00, 3.055261e-01)
(2.683177e+00, 3.213733e-01)
(2.649381e+00, 3.400294e-01)
(2.615017e+00, 3.523958e-01)
(2.580068e+00, 3.724403e-01)
(2.544517e+00, 3.878343e-01)
(2.508348e+00, 4.087773e-01)
(2.471541e+00, 4.300075e-01)
(2.434078e+00, 4.466976e-01)
(2.376609e+00, 4.736493e-01)
(2.277250e+00, 5.728975e-01)
(2.236189e+00, 6.040727e-01)
(2.215363e+00, 6.201954e-01)
(2.194337e+00, 6.436152e-01)
(2.183747e+00, 7.034985e-01)
(2.181623e+00, 7.091521e-01)
(2.179497e+00, 7.168291e-01)
(2.177369e+00, 7.253283e-01)
(2.175239e+00, 7.378446e-01)
(2.174173e+00, 7.404634e-01)
(2.174173e+00, 7.658948e-01)
(2.175239e+00, 7.743609e-01)
(2.177369e+00, 7.793755e-01)
(2.181623e+00, 7.865234e-01)
(2.179497e+00, 7.874278e-01)
(2.183747e+00, 7.936017e-01)
(2.194337e+00, 8.091876e-01)
(2.215363e+00, 8.284269e-01)
(2.236189e+00, 8.388544e-01)
(2.277250e+00, 8.580018e-01)
(2.376609e+00, 8.960805e-01)
(2.434078e+00, 9.099296e-01)
(2.471541e+00, 9.180719e-01)
(2.508348e+00, 9.247089e-01)
(2.544517e+00, 9.331577e-01)
(2.580068e+00, 9.358121e-01)
(2.615017e+00, 9.399287e-01)
(2.649381e+00, 9.441465e-01)
(2.683177e+00, 9.485353e-01)
(2.716420e+00, 9.541723e-01)
(2.749124e+00, 9.572766e-01)
(2.781304e+00, 9.594045e-01)
(2.812975e+00, 9.624443e-01)
(2.844148e+00, 9.627888e-01)
(2.874837e+00, 9.636979e-01)
(2.905053e+00, 9.659637e-01)
(2.934810e+00, 9.683411e-01)
(2.964117e+00, 9.692852e-01)
(2.992986e+00, 9.707292e-01)
(3.021428e+00, 9.736392e-01)
(3.049452e+00, 9.738487e-01)
(3.077069e+00, 9.752157e-01)
(3.104288e+00, 9.771619e-01)
(3.131117e+00, 9.777101e-01)
(3.157567e+00, 9.787061e-01)
(3.183645e+00, 9.796759e-01)
(3.234719e+00, 9.812993e-01)
(3.209360e+00, 9.815467e-01)
(3.356447e+00, 9.852366e-01)
(3.470352e+00, 9.875923e-01)
(3.577188e+00, 9.898664e-01)
(3.677604e+00, 9.924921e-01)
(3.772165e+00, 9.933042e-01)
(3.861365e+00, 9.945625e-01)
(3.945636e+00, 9.952101e-01)
(4.025361e+00, 9.963628e-01)
(4.100883e+00, 9.968301e-01)
(4.172502e+00, 9.975285e-01)
(4.305101e+00, 9.976850e-01)
(4.240494e+00, 9.978585e-01)
(4.366547e+00, 9.980765e-01)
(4.480739e+00, 9.985647e-01)
(4.425032e+00, 9.987318e-01)
(4.533837e+00, 9.989457e-01)
(4.584479e+00, 9.989820e-01)
(5.125722e+00, 1.000000e+00)
(5.242727e+00, 1.000000e+00)
(5.667603e+00, 1.000000e+00)
(5.486291e+00, 1.000000e+00)
( 6,  1)
};
%
\addplot [color = blue, blue, thick, dashed] coordinates {(2.7164, 0) (2.7164, 1)};
\addplot [color = red, red] coordinates {(2.993, 0) (2.993, 1)};
\end{axis}
\draw [rounded corners,fill=white] (0.05+1.65,1.7+0.7) rectangle (1.35+1.75,0.6+0.7);
\draw [red] (0.1+1.65,1.5+0.7) -- (0.35+1.65,1.5+0.7);
\node [right] at (0.28+1.65,1.5+0.7) {\scriptsize{U.B.~on}};
\node [right] at (0.28+1.65,1.5-0.3+0.7) {\scriptsize{MAP th.}};
\draw [blue, thick, dashed] (0.1+1.65,1.5-0.65+0.7) -- (0.35+1.65,1.5-0.65+0.7);
\node [right] at (0.28+1.65,1.5-0.65+0.7) {\scriptsize{MAP th.}};
\node [below] at (1.5,-0.3) {\footnotesize{(h) $\mathcal{S}_8$, $64$-QAM, Natural}};
\end{tikzpicture}
\end{center}
\vspace{-0.5cm}
\caption{EBP-GEXIT chart of the systems $\bm{\mathcal{S}_1}$ to $\bm{\mathcal{S}_8}$ for various modulators}
\label{Figure_combined_EBP_GEXIT}
\end{figure*}

Note that 
$\mathcal{S}_1$ to $\mathcal{S}_4$ are SC-TC and $\mathcal{S}_5$ to $\mathcal{S}_8$ are LDPC codes.
For $\mathcal{S}_4$ and $\mathcal{S}_8$ we have chosen non-Gray mapping.
%
The obtained approximate EBP-GEXIT charts of all the above mentioned schemes are provided in Fig.~\ref{Figure_combined_EBP_GEXIT}. For EBP-GEXIT charts, on the X-axis we have the channel entropy and on the Y-axis we have the corresponding EBP-GEXIT function.
Tables~\ref{Table_MAP_estimates_Serially_concat} and \ref{Table_MAP_estimates_LDPC} summarize the obtained thresholds.
%
The MAP thresholds are estimated by applying the Maxwell construction~\cite{Maxell_Urbanke_2008} to the approximate EBP-GEXIT charts illustrated in Fig.~\ref{Figure_combined_EBP_GEXIT} (refer \cite[Sec.~3.20]{Urbanke_Richardson_MCT_Book},\cite{Maxell_Urbanke_2008} for details about the Maxwell construction). An upper bound (U.B.) on the MAP threshold is obtained by applying area theorem to the respective EBP-GEXIT charts (see \cref{Theorem_area_theorem_nonbin_AWGN_our} and \cite[Theorem~5]{GAT_Urbanke_2009}). Spatial-coupling of LDPC codes with Gray mapping is studied in \cite{Yedla_LDPC_BICM_arxiv}. While the BP-threshold of the spatially-coupled $\mathcal{S}_5$ system with $L=64$ provided in \cite{Yedla_LDPC_BICM_arxiv} is  $E_b/N_0 = 0.54$dB ($h = 0.473$), our estimated MAP threshold is $E_b/N_0 = 0.601$dB ($h = 0.4682$) (see \cref{Table_MAP_estimates_LDPC} and for QPSK we have $E_b/N_0 = E_s/N_0$). The small difference in the two values
might be due to various approximations involved while computing our EBP-GEXIT chart (e.g.~polynomial approximations for the constituent EXIT charts, Gauss-Hermite procedure etc).

%
For spatially-coupled SC-TC with Gray mapping, we also provide BP threshold of their respective spatially-coupled versions in \cref{Table_MAP_estimates_Serially_concat}\footnote{
We focus on the spatial coupling of SC-TC with Gray mapping since for 
other systems the decoding of SC-TC or LDPC codes becomes doubly iterative~\cite{ten2003design}. We plan to do spatial coupling analysis of the these systems in future work.}.
For obtaining the BP threshold of spatially-coupled SC-TC, we have chosen $B = [1/2, 1/2]$ ($m_s=1$) and $L=200$.
Concerning the asymptotic EXIT convergence criterion, we choose that if, given a channel parameter, after $10^5$ BP iterations, the \emph{a posteriori} MI corresponding to the different stages does not converge to $1$, then the decoder has failed to recover the transmitted bits.
For comparison, in \cref{Table_MAP_estimates_Serially_concat} we also include the threshold bound given by the EXIT chart area theorem, where the EXIT area is computed for the combined detector and inner code. component~\cite{hagenauer2004exit}.
%
From \cref{Table_MAP_estimates_Serially_concat} it can be seen that, the BP threshold of spatially-coupled SC-TC is close to the MAP threshold estimated from the EBP-GEXIT chart.
%
%
%
The examples where the estimated MAP threshold is away from their corresponding bounds indicate a suboptimal choice for the outer code. 
%
%
The small difference between the MAP threshold estimated via the EBP-GEXIT chart and that of BP threshold of spatially-coupled SC-TC might be due to 
various approximations involved while computing the EBP-GEXIT chart.
%
%
Due to these reasons, we conjecture that the BP threshold of spatially-coupled SC-TC converges to the MAP threshold estimated using the EBP-GEXIT chart.

%
%
%
%
%

\begin{table}[htbp]

\begin{center}

\begin{tabular}{|c|c|c|c|c|c|c|c|c|c|}
\hline
  \multirow{3}{*}{scheme} 
& \multicolumn{2}{c|}{MAP threshold} 
& \multicolumn{3}{c|}{BP threshold} 
& \multicolumn{2}{c|}{U.B.~on MAP th.} 
& \multicolumn{2}{c|}{EXIT area} \\
& \multicolumn{2}{c|}{via EBP-GEXIT} 
& \multicolumn{3}{c|}{of SC-TC} 
& \multicolumn{2}{c|}{via EBP-GEXIT}
& \multicolumn{2}{c|}{} \\
\cline{2-10}
& $h$ & $E_s/N_0$ (dB) 
& $h$ & $E_s/N_0$ (dB) & Rate 
& $h$ & $E_s/N_0$ (dB) 
& $h$ & $E_s/N_0$ (dB)  \\
\hline
$\mathcal{S}_1$ 
& $0.4893$ & $-2.71$
& $0.4793$ &  $-2.55$  & $0.4975$
& $0.4974$ &  $-2.79$
& $0.4963$ &  $-2.74$\\
\hline
$\mathcal{S}_2$ 
& $0.7448$ & $4.02$
& $0.7435$ &  $4.05$  & $0.2488$
& $0.7456$ &  $4.00$
& $0.7533$ &  $3.80$\\
\hline
$\mathcal{S}_3$ 
& $0.7241$ &  $1.20$
& $0.7211$ &  $1.27$  & $0.2488$
& $0.7315$ &  $1.03$
& $0.7433$ &  $0.73$\\
\hline
%
\end{tabular}

\end{center}

\vspace{-0.5cm}
\caption{Estimates of MAP thresholds for various SC-TC with Gray mapping}
\label{Table_MAP_estimates_Serially_concat}
\end{table}

%

\begin{table}[htbp]

\begin{center}

\begin{tabular}{|c|c|c|c|c|}
\hline
\multirow{2}{*}{scheme} 
& \multicolumn{2}{c|}{MAP th.~via EBP-GEXIT} 
& \multicolumn{2}{c|}{U.B.~on MAP th.~via EBP-GEXIT} \\
\cline{2-5}
& $h$ 
& $E_s/N_0$ (dB) 
& $h$ 
& $E_s/N_0$ (dB) \\
\hline
%
%
$\mathcal{S}_4$ & $2.4096$ & $3.2393$ & $2.9874$ & $0.1412$ \\
\hline 
%
%
$\mathcal{S}_5$ & $0.4682$ & $0.601$ & $0.4949$ & $0.253$ \\
\hline
%
%
$\mathcal{S}_6$ & $0.5123$ & $0.022$ & $0.5238$ &  $-0.133$ \\
\hline
%
%
$\mathcal{S}_7$ & $0.514$ & $-3.01$ & $0.514$ & $-3.01$ \\
\hline
$\mathcal{S}_8$ & $2.7164$ & $10.0549$ & $2.993$ & $9.0319$  \\
\hline
\end{tabular}

\end{center}

\vspace{-0.5cm}
\caption{Estimates of MAP thresholds for various SC-TC and LDPC codes}
\label{Table_MAP_estimates_LDPC}

\end{table}

%

It is important to mention that for the serially concatenated systems with a non-Gray mapper as an inner code and a convolutional code as an outer code, the BP threshold does not exist. From an EXIT chart point of view, this can be inferred by the fact that the EXIT curves of the constituent codes intersect before reaching the point $(1,1)$. A similar behaviour is also exhibited by low-density generator-matrix (LDGM) codes. Therefore, our method or EBP-GEXIT in general is not applicable for the MAP threshold estimation of these schemes. This is a well known problem and different methods are proposed to tackle this limitation (e.g. \cite{kumar2014threshold} studies this via potential threshold approach).
Note that for the system $\mathcal{S}_4$ considered in our work, we have chosen a SC-TC system with two convolutional codes as inner and outer codes respectively and a non-Gray mapping is chosen for modulation. For such a SC-TC system, the combined EXIT curve of the detector and inner convolutional code is considered and the BP and MAP thresholds are well defined for such setups (See Fig.~\ref{Figure_combined_EBP_GEXIT}-(d)).

%


\section{Conclusions and future work}
\label{Section_Conclusion}
We studied the problem of estimating the MAP threshold for LDPC/GLDPC/DGLDPC codes and SC-TC families, when the transmission is over non-binary complex-input AWGN channel.
We extended the existing results to obtain the GEXIT function over complex AWGN channel and provided a tractable method for fast evaluation of an approximate EBP-GEXIT chart, based on the Gaussian approximation.
We estimated the MAP thresholds for various families of LDPC/GLDPC/DGLDPC codes and SC-TC for Gray and non-Gray mappings. For SC-TC system with Gray mapping, we also studied the threshold saturation phenomenon.
Numerical results indicate that the BP threshold of spatially-coupled SC-TC does saturate to the MAP threshold obtained using the EBP-GEXIT chart.


Since our proposed method for the computation of EBP-GEXIT charts only requires the knowledge of the constituent EXIT charts, this opens up the applicability of our method to a variety of setups such as multiple-input multiple-output (MIMO) system, intersymbol interference (ISI) channel, and Rayleigh fading channel. 
As a part of the future work, it will also be interesting to investigate schemes such as LDGM codes for which BP threshold does not exist,
under the light of our proposed framework.
%
%
%
%


%
\section*{Appendix~A: GEXIT function for non-binary complex-input AWGN channel}
%

The key idea of the proof of \cref{Theorem_GEXIT_nonbin_complex_AWGN_our} comes from Lemma~1 of \cite{Pfister_ISI_2012}.
This lemma provides an expression for the $t$-th GEXIT function $g_t(h)$ for non-binary real-input AWGN channel.\\
%
%
%
\textit{Lemma~1 of \cite{Pfister_ISI_2012}:}
%
Consider $f_{t,\xi}(\phi)$ and $\phi_{[\xi]}$ defined in  
\cref{subsection_GEXIT_non_Gray}.
%
Let $p(\xi) = \mathbb{P}[X_t = \xi]$, $p(y_t|\xi^{\prime}) = \mathbb{P}[Y_t = y_t|X_t = \xi^{\prime}]$, and $p^{\prime}(y_t|\xi) = \frac{\partial}{\partial \epsilon}p(y_t|\xi)$, where $\epsilon = \frac{-1}{2 \sigma^2}$.
%
Then $g_t(h)$ for $|\mathbb{X}|$-ary real-input AWGN channel is given by,
\begin{align*}
\mbox{~~~~~~~~~~~}
g_t(h) = \sum_{\xi \in \mathbb{X}} p(\xi) \int_{\phi} f_{t,\xi}(\phi) %
\frac{\int_{-\infty}^{\infty} p^{\prime}(y_t|\xi) \log_2 \left\{ \frac{\sum_{\xi^{\prime}} \phi_{[\xi^{\prime}]} p(y_t|\xi^{\prime})}{\phi_{[\xi]} p(y_t|\xi)} \right\} \mathrm{d}y_t}
{\int_{-\infty}^{\infty} \sum_{\xi}p(\xi) p^{\prime}(y_t|\xi) \log_2 \left\{ \frac{\sum_{\xi^{\prime}} p(\xi^{\prime}) p(y_t|\xi^{\prime})}{p(\xi) p(y_t|\xi)} \right\} \mathrm{d}y_t} \mathrm{d}\phi. \mbox{~~~~~~~~~~~}\square
%
%
\end{align*}
%

It can be easily verified that this lemma hold true for complex-input AWGN channel as well and hence we use it to obtain the required expression for the theorem.
Towards this we next obtain an expression for $p^{\prime}(y_t|\xi)$ and 
$p(y_t|\xi^{\prime})/p(y_t|\xi)$ for complex-input AWGN channel.
%
%
%
We first obtain an expression for $p^{\prime}(y_t|\xi)$. Since we have assumed that the distribution of the noise corresponding to both real and imaginary parts is $\mathcal{N}(0,\sigma^2)$, the distribution of $Y_t$ under the condition $\xi$ is bivariate Gaussian. Using this we get,
\begin{equation}
\begin{aligned}
p^{\prime}(y_t|\xi) &= \frac{\partial}{\partial \epsilon} p(y_t|\xi) 
= \frac{\partial}{\partial \epsilon} \frac{1}{2 \pi \sigma^2} e^{-\frac{|y_t-\xi|^2}{2 \sigma^2}}
\stackrel{(a)}{=} \frac{\partial \sigma}{\partial \epsilon} \Bigg(\frac{\partial}{\partial \sigma} \frac{1}{2 \pi \sigma^2} e^{-\frac{|y_t-\xi|^2}{2 \sigma^2}} \Bigg),\\
%
%
%
&\stackrel{(b)}{=} p(y_t|\xi) \left( |y_t-\xi|^2 - 2\sigma^2\right),
\end{aligned}
\label{Eqn_derivative_complex_AWGN_part_1}
\end{equation}
where the equality in $(a)$ follows from the chain rule of derivative. Since $\epsilon = \frac{-1}{2 \sigma^2}$, $\sigma$ is a function of $\epsilon$ and by solving the derivative we get 
$\frac{\partial \sigma}{\partial \epsilon} = \sigma^3$. 
Substituting this in $(a)$ and solving the derivative we obtain $(b)$.
%
The fraction $p(y_t|\xi^{\prime})/p(y_t|\xi)$ is given by
\begin{align}
\frac{p(y_t|\xi^{\prime})}{p(y_t|\xi)} &= \left(\frac{1}{2 \pi \sigma^2} e^{-\frac{|y_t-\xi^{\prime}|^2}{2 \sigma^2}} \right) \Bigg/ \left(\frac{1}{2 \pi \sigma^2} e^{-\frac{|y_t-\xi|^2}{2 \sigma^2}}\right)
%
%
%
= \exp \left[ \frac{|y_t-\xi|^2 - |y_t-\xi^{\prime}|^2}{2 \sigma^2}  \right].
\label{Eqn_AWGN_complex_ratio1}
\end{align}
%
%
Since in general we have $N >> |\mathbb{X}|$, in presence of an ideal interleaver we can assume that the transmitted modulated symbols are equally likely, i.e., $p(\xi) = 1/|\mathbb{X}|$ for any $\xi \in \mathbb{X}$. Using this and substituting
\cref{Eqn_derivative_complex_AWGN_part_1} and \cref{Eqn_AWGN_complex_ratio1} in Lemma~1 of \cite{Pfister_ISI_2012} we get the required expression of the theorem.
This completes the proof of the theorem.
\hfill $\blacksquare$

%
\section*{Appendix~B: Proof of \cref{Theorem_area_theorem_nonbin_AWGN_our} (Area theorem for the system of \Fig~\ref{Figure_general_digi_comm_system})}
The proof of \cref{Theorem_area_theorem_nonbin_AWGN_our} follows directly by applying the generalized area theorem (GAT)~\cite[Theorem~1]{GAT_Urbanke_2009}. 
While in \cite{GAT_Urbanke_2009}, GAT was derived for BMS, we observe that this GAT is also applicable for any $|\mathbb{X}|$-ary, complex-input memoryless channel.
Suppose the channel input symbols $\mathbf{X} = [ X_1 \mbox{~} X_2 \mbox{~} \ldots \mbox{~} X_N]$ are transmitted via the set of parallel independent memoryless channels parameterized by $h_1, h_2, \ldots, h_N$ respectively to receive $\mathbf{Y} = [ Y_1 \mbox{~} Y_2 \mbox{~} \ldots \mbox{~} Y_N]$. Then from GAT, 
\begin{align*}
%
\mathrm{d}
H(\mathbf{X}|\mathbf{Y}) = \sum_{t=1}^N \frac{\partial H(X_t|\mathbf{Y})}{\partial h_t} \mathrm{d}h_t.
\end{align*}
If all the individual channel parameters $h_1, h_2, \ldots, h_N$ in are parameterized in a smooth way by a common parameter $h$, then the GEXIT function $g(h)$ is defined as~\cite{GAT_Urbanke_2009}
\begin{align*}
g(h) = \sum_{t=1}^N \frac{\partial H(X_t|\mathbf{Y})}{\partial h_t} \frac{\mathrm{d}h_t}{\mathrm{d}h}\Big|_{h}.
%
%
\end{align*}
%

Each $Y_t$ is a function of the $t$-th channel parameter $h_t$ and can be denoted by $Y_t(h_t)$. 
Integrating $g(h)$ from $0$ to $|\mathbb{X}|$ we get (refer to the discussion after Definition~3 of \cite{GAT_Urbanke_2009}),
\begin{equation}
\begin{aligned}
 \int_{\underline{h}}^{\bar{h}} g(h) = \frac{1}{N} \Big[ H(\mathbf{X}|\mathbf{Y}(|\mathbb{X}|)) - H(\mathbf{X}|\mathbf{Y}(0)) \Big]
\stackrel{(a)}{=} \frac{1}{N} \big[ k - 0 \big]
%
\stackrel{(b)}{=} \frac{km}{n}.
%
%
\end{aligned}
\end{equation}
The equality in $(a)$ is obtained since the entropy $H(\mathbf{X}|\mathbf{Y}(0))$, which is the uncertainty about $\mathbf{X}$ in presence of zero noise, is equal to $0$. 
Note that for the AWGN channel, noise entropy $h = |\mathbb{X}|$ correspond to large enough (ideally infinite) noise variance such that the received $\mathbf{Y}$ does not provide any information about the transmitted $\mathbf{X}$. This implies that $H(\mathbf{X}|\mathbf{Y}(|\mathbb{X}|))= H(\mathbf{X})$.
Since modulation scheme does not change the entropy of the transmitted codewords we have $H(\mathbf{X}) = k$, where $k$ is the dimension of the code.
The equality in $(b)$ is obtained since for $m$-ary modulation scheme we have $N=n/m$ (see \cref{Section_System_Model}) and this completes the proof.
\hfill $\blacksquare$
%

\section*{Appendix C: Approximation for the GEXIT kernel of BAWGN channel} 

%
%
%
%
%
%
%
%
%
%
%
%
%
%
%

For a random variable $W \sim \mathcal{N}(2/\sigma^2, 4/\sigma^2)$ and real number $z$, define a function $f(h,z)$ as 
%
\begin{equation}
\begin{aligned}
f(h,z) &= \mathbb{E}_W \left[\frac{1}{1+e^{W+z}} \right] 
= \int_{-\infty}^{\infty} \frac{1}{\sqrt{2\pi (4/\sigma^2)}} \frac{e^{- \frac{(w-(2/\sigma^2))^2}{8/ \sigma^2}}}{1+e^{w+z}} \mathrm{d}w,
\end{aligned}
\label{Eqn_BAWGN_kernel_generic_Nr_Dr}
\end{equation}
where $h= 1-J(2/\sigma)$. Using this in \cref{Eqn_BAWGN_kernel} we have $l^{c_{BAWGN(h)}}(z) = f(h,z)/f(h,z=0)$.
The function $f(h,z)$ can be approximated using Marquardt-Levenberg algorithm~\cite{Marquardt_Levenberg_Book} as follows.
\begin{equation}
\scriptsize
f(\sigma,z) \approx
	      \begin{cases}
	      1       									& \quad 	\text{if } z \leq L(h)	\\
	      1-e^{A_3(h)z^3 + A_2(h)z^2 + A_1(h)z + A_0(h)}      	& \quad 	\text{if } L(h) < z < M(h)	\\		
	      0 									& \quad 	\text{if } z \geq M(h)
	      \end{cases}
\end{equation}
where $L(h), M(h), A_0(h), A_1(h), A_2(h),$ and $A_3(h)$ are approximated as polynomials of degree $10$ and are given by
\scriptsize{
\begin{align*}
L(h)&= -92218h^{10}+ 490818h^{9} -1127499h^{8}+ 1463798h^{7} -1181473h^{6}+ 614716h^{5} -207094h^{4}+ 44333h^{3} -5817h^{2}+  467h  -38 \\
M(h)&= -33578h^{10}+ 175895h^{9} -397298h^{8}+ 506819h^{7} -401852h^{6}+ 205453h^{5} -68054h^{4}+ 14322h^{3} -1837h^{2}+  136h+    1\\
A_0(h)&= 117.76h^{10} -610.96h^{9}+ 1344.72h^{8} -1634.10h^{7}+ 1195.44h^{6} -538.70h^{5}+ 146.66h^{4} -22.38h^{3}+ 1.19h^{2} -0.33h\\
A_1(h)&= 28.89h^{10} -155.08h^{9}+ 369.42h^{8} -512.31h^{7}+ 453.43h^{6} -262.89h^{5}+ 98.69h^{4} -22.90h^{3}+ 3.18h^{2}+ 0.01h\\
A_2(h)&= 10.94h^{10} -62.92h^{9}+ 154.13h^{8} -210.31h^{7}+ 175.60h^{6} -93.03h^{5}+ 31.50h^{4} -6.78h^{3}+ 0.88h^{2} -0.11h\\
A_3(h)&= 0.47h^{10} -1.58h^{9}+ 1.47h^{8}+ 0.85h^{7} -2.78h^{6}+ 2.44h^{5} -1.11h^{4}+ 0.29h^{3} -0.04h^{2}+ 0.01h.
\end{align*}
}

\bibliographystyle{IEEEtran}
\bibliography{LDPC}

\begin{thebibliography}{10}
\providecommand{\url}[1]{#1}
\csname url@samestyle\endcsname
\providecommand{\newblock}{\relax}
\providecommand{\bibinfo}[2]{#2}
\providecommand{\BIBentrySTDinterwordspacing}{\spaceskip=0pt\relax}
\providecommand{\BIBentryALTinterwordstretchfactor}{4}
\providecommand{\BIBentryALTinterwordspacing}{\spaceskip=\fontdimen2\font plus
\BIBentryALTinterwordstretchfactor\fontdimen3\font minus
  \fontdimen4\font\relax}
\providecommand{\BIBforeignlanguage}[2]{{%
\expandafter\ifx\csname l@#1\endcsname\relax
\typeout{** WARNING: IEEEtran.bst: No hyphenation pattern has been}%
\typeout{** loaded for the language `#1'. Using the pattern for}%
\typeout{** the default language instead.}%
\else
\language=\csname l@#1\endcsname
\fi
#2}}
\providecommand{\BIBdecl}{\relax}
\BIBdecl

\bibitem{Urbanke_Richardson_MCT_Book}
T.~Richardson and R.~Urbanke, \emph{Modern Coding Theory}.\hskip 1em plus 0.5em
  minus 0.4em\relax Cambridge University Press, 2008.

\bibitem{GAT_Urbanke_2009}
C.~M{\'e}asson, A.~Montanari, T.~Richardson, and R.~Urbanke, ``The generalized
  area theorem and some of its consequences,'' \emph{IEEE Trans.~on
  Info.~Theory}, vol.~55, no.~11, pp. 4793--4821, Nov. 2009.

\bibitem{Maxell_Urbanke_2008}
C.~M{\'e}asson, A.~Montanari, and R.~Urbanke, ``Maxwell construction: The
  hidden bridge between iterative and maximum a posteriori decoding,''
  \emph{IEEE Trans.~on Info.~Theory}, vol.~54, no.~12, pp. 5277--5307, Dec.
  2008.

\bibitem{Kudekar_ThresholdSaturationIT}
S.~Kudekar, T.~Richardson, and R.~Urbanke, ``Threshold saturation via spatial
  coupling: Why convolutional {LDPC} ensembles perform so well over the
  {BEC},'' \emph{IEEE Transactions on Information Theory}, vol.~57, no.~2, pp.
  803--834, Feb. 2011.

\bibitem{Yedla_potential_func_2014}
A.~{Yedla}, Y.~{Jian}, P.~{Nguyen}, and H.~{Pfister}, ``A simple proof of
  maxwell saturation for coupled scalar recursions,'' \emph{IEEE Transactions
  on Information Theory}, vol.~60, no.~11, pp. 6943--6965, November 2014.

\bibitem{kumar2014threshold}
S.~Kumar, A.~J. Young, N.~Macris, and H.~D. Pfister, ``Threshold saturation for
  spatially coupled {LDPC} and {LDGM} codes on {BMS} channels,'' \emph{IEEE
  Trans. on Inf. Theory}, vol.~60, no.~12, pp. 7389--7415, 2014.

\bibitem{moloudi2016spatially}
S.~Moloudi, M.~Lentmaier, and A.~G. i~Amat, ``Spatially coupled turbo-like
  codes,'' \emph{IEEE Trans.~on~Info.~Theory}, vol.~63, no.~10, pp. 6199--6215,
  Oct 2017.

\bibitem{SCTC_AWGN_TCOM_2019}
S.~{Moloudi}, M.~{Lentmaier}, and A.~{Graell i Amat}, ``Spatially coupled
  turbo-like codes: A new trade-off between waterfall and error floor,''
  \emph{IEEE Transactions on Communications}, vol.~67, no.~5, pp. 3144--3123,
  2019.

\bibitem{Tarik_Globecomm_2017}
T.~Benaddi, C.~Poulliat, and R.~Tajan, ``A general framework and optimization
  for spatially-coupled serially concatenated systems,'' in \emph{Proc.~of IEEE
  GLOBECOM}, Singapore, December 2017, pp. 1--6.

\bibitem{Yedla_LDPC_BICM_arxiv}
A.~{Yedla}, M.~{El-Khamy}, J.{Lee}, and I.~{Kang}, ``Performance of
  spatially-coupled {LDPC} codes and threshold saturation over {BICM}
  channels,'' \emph{arXiv preprint arXiv:1303.0296[cs.IT]}, 2013.

\bibitem{Our_ISIT_2018}
A.~Yardi, I.~Andriyanova, and C.~Poulliat, ``{EBP-GEXIT} charts over the
  binary-input {AWGN} channel for generalized and doubly-generalized {LDPC}
  codes,'' in \emph{Proc.~of IEEE ISIT}, Vail, Colorado, USA, June 2018, pp.
  496--500.

\bibitem{Our_ISIT_2019}
T.~{Benaddi}, A.~{Yardi}, C.~{Poulliat}, and I.~{Andriyanova}, ``Estimating the
  maximum a posteriori threshold for serially concatenated turbo codes,'' in
  \emph{Proc.~of IEEE ISIT}, Paris, France, July 2019, pp. 1347--1351.

\bibitem{BICM_Fabregas_Book}
A.~G. i~F\`{a}bregas, A.~Martinez, and G.~Caire, ``Bit-interleaved coded
  modulation,'' \emph{Foundations and Trends in Communications and Information
  Theory}, vol.~5, no. 1-2, pp. 1--153, 2008.

\bibitem{Ryan_Costello_Book}
W.~Ryan and S.~Lin, \emph{Channel codes: classical and modern}.\hskip 1em plus
  0.5em minus 0.4em\relax Cambridge University Press, 2009.

\bibitem{Brink_EXIT_2004}
A.~Ashikmin, G.~Kramer, and S.~ten Brink, ``Extrinsic information transfer
  functions: Model and erasure channel properties,'' \emph{IEEE Trans.~on
  Info.~Theory}, vol.~50, no.~11, pp. 2657--2673, Nov. 2004.

\bibitem{Bennatan_Non_binary}
A.~Bennatan and D.~Burshtein, ``Design and analysis of nonbinary {LDPC} codes
  for arbitrary discrete-memoryless channels,'' \emph{IEEE Transactions on
  Information Theory}, vol.~52, no.~2, pp. 549--583, 2006.

\bibitem{Pfister_ISI_2012}
P.~Nguyen, A.~Yedla, and H.~P. adn K.~Narayanan, ``Threshold saturation of
  spatially-coupled codes on intersymbol-interference channels,'' in
  \emph{Proceedings of IEEE ICC}, Ottawa, Canada, June 2012, pp. 2181--2186.

\bibitem{BICM_Szczecinski_book}
L.~Szczeci{\'n}ski and A.~Alvarado, \emph{Bit-Interleaved Coded
  Modulation:Fundamentals, Analysis, and Design}.\hskip 1em plus 0.5em minus
  0.4em\relax Chichester, United Kingdom: John Wiley and Sons, 2015.

\bibitem{i2008bit}
A.~G. i~F{\`a}bregas, A.~Martinez, G.~Caire \emph{et~al.}, ``Bit-interleaved
  coded modulation,'' \emph{Foundations and Trends{\textregistered} in
  Communications and Information Theory}, vol.~5, no. 1--2, pp. 1--153, 2008.

\bibitem{Caire_GA}
A.~F\`{a}bregas, A.~Martinez, and G.~Caire, ``Error probability of
  bit-interleaved coded modulation using the {G}aussian approximation,'' in
  \emph{Proc.~of Conference on Information Science and Systems}, New Jersey,
  USA, March 2004.

\bibitem{Szczecinski_PSK}
L.~Szczeci{\'n}ski and M.~Benjillali, ``Probability density functions of
  logarithmic likelihood ratios in phase shift keying {BICM},'' in
  \emph{Proceedings of IEEE GLOBECOM}, San Francisco, USA, November 2006, pp.
  1--6.

\bibitem{Szczecinski_QAM}
M.~Benjillali, L.~Szczeci{\'n}ski, and S.~A\"{i}ssa, ``Probability density
  functions of logarithmic likelihood ratios in rectangular {QAM},'' in
  \emph{Proceedings of 23rd Biennial Symposium on Communications}, 2006.

\bibitem{WiMax_standard_Book}
\emph{IEEE 802.16m Evaluation Methodology Document}.\hskip 1em plus 0.5em minus
  0.4em\relax IEEE 802.16 Broadband Wireless Access Working Group, 2008.

\bibitem{Liva_Quasi_Cyclic_2008}
G.~Liva, W.~Ryan, and M.~Chiani, ``Quasi-cyclic generalized {LDPC} codes with
  low error floors,'' \emph{IEEE Transactions on Communications}, vol.~56,
  no.~1, pp. 49--57, January 2008.

\bibitem{Yige_DG_ISIT_2006}
Y.~Wang and M.~Fossorier, ``Doubly generalized {LDPC} codes,'' in
  \emph{Proceedings of IEEE International Symposium on Information Theory},
  Seattle, WA, USA, July 2006, pp. 669--673.

\bibitem{GA_Urbanke_2001}
S.~Chung, T.~Richardson, and R.~Urbanke, ``Analysis of sum-product decoding of
  {LDPC} codes using a {G}aussian approximation,'' \emph{IEEE Transactions on
  Information Theory}, vol.~47, no.~2, pp. 657--670, Feb. 2001.

\bibitem{Gauss_Quadrature_book}
M.~Abramowitz and I.~Stegun, \emph{Handbook of Mathematical Functions}.\hskip
  1em plus 0.5em minus 0.4em\relax New York, USA: Dover, 1970.

\bibitem{GEXIT_Turbo_Urbanke_2005}
C.~M{\'e}asson, R.~Urbanke, A.~Montanari, and T.~Richardson, ``Maximum a
  posteriori decoding and turbo codes for general memoryless channels,'' in
  \emph{Proc.~of ISIT}, Australia, Sept. 2005, pp. 1241--1245.

\bibitem{ten2001convergence}
S.~ten Brink, ``Convergence behavior of iteratively decoded parallel
  concatenated codes,'' \emph{IEEE Trans.~on~Commun.}, vol.~49, no.~10, pp.
  1727--1737, 2001.

\bibitem{ten2003design}
S.~ten Brink and G.~Kramer, ``Design of repeat-accumulate codes for iterative
  detection and decoding,'' \emph{IEEE Transactions on Signal Processing},
  vol.~51, no.~11, pp. 2764--2772, 2003.

\bibitem{Brink_Modulation_2004}
S.~ten Brink, G.~Kramer, and A.~Ashikmin, ``Design of low-density parity-check
  codes for modulation and detection,'' \emph{IEEE Transactions on
  Communications}, vol.~52, no.~4, pp. 670--678, April 2004.

\bibitem{Gauss_quadrature_multivariate_Jackel}
P.~J{\"a}ckel, ``A note on multivariate gauss-hermite quadrature,''
  \emph{Oxford University, Technical Report}, May, 2005.

\bibitem{costello2016new}
D.~Costello, M.~Lentmaier, and D.~Mitchell, ``New perspectives on braided
  convolutional codes,'' in \emph{Proc.~of ISTC}, 2016, pp. 400--405.

\bibitem{moloudi2014spatially}
S.~Moloudi, M.~Lentmaier, and A.~G. i~Amat, ``Spatially coupled turbo codes,''
  in \emph{Proc.~of IEEE ISTC}, 2014, pp. 82--86.

\bibitem{cammerer2016triggering}
S.~Cammerer, V.~Aref, L.~Schmalen, and S.~ten Brink, ``Triggering wave-like
  convergence of tail-biting spatially coupled {LDPC} codes,'' in \emph{Annual
  Conference on Information Science and Systems (CISS)}.\hskip 1em plus 0.5em
  minus 0.4em\relax IEEE, 2016, pp. 93--98.

\bibitem{tazoe2012efficient}
K.~Tazoe, K.~Kasai, and K.~Sakaniwa, ``Efficient termination of
  spatially-coupled codes,'' in \emph{ITW}.\hskip 1em plus 0.5em minus
  0.4em\relax IEEE, 2012, pp. 30--34.

\bibitem{Yige_DG_Globecomm_2006}
Y.~Wang and M.~Fossorier, ``{EXIT} chart analysis for doubly generalized {LDPC}
  codes,'' in \emph{Proceedings of IEEE Globecomm}, San Francisco, California,
  USA, November 2006, pp. 1--6.

\bibitem{hagenauer2004exit}
J.~Hagenauer, ``The exit chart-introduction to extrinsic information transfer
  in iterative processing,'' in \emph{Proc. 12th European Signal Processing
  Conference (EUSIPCO)}, 2004, pp. 1541--1548.

\bibitem{Marquardt_Levenberg_Book}
W.~Press, S.~Teukolsky, W.~Vetterling, and B.~Flannery, \emph{Numerical recipes
  in C}.\hskip 1em plus 0.5em minus 0.4em\relax New York, USA: Cambridge
  University Press, 1997.

\end{thebibliography}

\end{document}